\documentclass[10pt,tightenlines,showpacs,letterpaper,aps,prd,longbibliography,twocolumn,nofootinbib,superscriptaddress,tikz]{revtex4-1}

\usepackage[nice]{nicefrac}
\usepackage{array}
\usepackage{amsmath}
\usepackage{float}
\usepackage{graphicx}
\usepackage{enumerate,amsthm,amssymb,color}
\usepackage{bbm}
\usepackage{bm}
\usepackage[french, english]{babel}
\usepackage{synttree}
\usepackage{caption}
\usepackage{subfig}
\usepackage{multirow}
\usepackage[svgnames]{xcolor}
\graphicspath{ {Figures/} }
\usepackage{hyperref}
\usepackage[latin1]{inputenc}
\usepackage{tikz}
\usepackage{pgfplots}
\usetikzlibrary{decorations.pathreplacing}
\usepackage{float}
\usepackage{tkz-euclide}
\pgfplotsset{compat=newest}
\usepgfplotslibrary{fillbetween}
\usetikzlibrary{patterns}

\newcommand{\COMMENT}[1]{}
\newcommand{\guess}{\operatorname{guess}}
\newcommand{\Models}{\operatorname{Models}}
\newcommand{\prep}{\operatorname{prep}}

\newcommand{\blk}{\color{black}}

\newtheorem{theorem}{Theorem}

\newtheorem{lemma}{Lemma}
\newtheorem*{conjecture*}{Conjecture}
\newtheorem*{theorem*}{Theorem}
\newtheorem*{lemma*}{Lemma}

\newtheorem{manualtheoreminner}{Lemma}
\newenvironment{manualtheorem}[1]{%
	\renewcommand\themanualtheoreminner{#1}%
	\manualtheoreminner
}{\endmanualtheoreminner}

\begin{document}

\title{Alternative robust ways of witnessing nonclassicality in the simplest scenario}

\author{Massy Khoshbin}\email{massy@math.ucsb.edu}\affiliation{Department of Mathematics, University of California, Santa Barbara, CA 93106, USA}

\author{Lorenzo Catani}\email{lorenzo.catani4@gmail.com}\affiliation{INL -- International Iberian Nanotechnology Laboratory, Braga, Portugal} 

\author{Matthew Leifer} \affiliation{Institute for Quantum Studies and Schmid College of Science and Technology, Chapman University, One University Drive, Orange, CA 92866, USA}


\begin{abstract}
	
	
	
	In this paper we relate notions of nonclassicality in what is known as the simplest nontrivial scenario (a prepare and measure scenario composed of four preparations and two binary-outcome tomographically complete measurements). Specifically, we relate the established method developed in [Pusey, PRA 98,022112(2018)] to witness a violation of preparation noncontextuality, that is not suitable in experiments where the operational equivalences to be tested are specified in advance, with an approach based on the notion of bounded ontological distinctness for preparations, defined in [Chaturvedi and Saha, Quantum 4, 345 (2020)].  
In our approach, we test bounded ontological distinctness for 
two particular preparations that are relevant in certain information processing tasks in that they are associated with the even and odd parity of the bits to communicate. When there exists an ontological model where this distance is preserved we talk of \textit{parity preservation}.  Our main result provides a noise threshold under which violating parity preservation (and so bounded ontological distinctness) agrees with the established method for witnessing preparation contextuality in the simplest nontrivial scenario. This is achieved by first relating the violation of parity preservation to the quantification of contextuality in terms of inaccessible information as developed in [Marvian, arXiv:2003.05984(2020)], that 
 we also show, given the way we quantify noise, to be more robust in witnessing contextuality 
	than Pusey's noncontextuality inequality. As an application of our findings, we treat the case of two-bit parity-oblivious multiplexing in the presence of noise. 
	 In particular, given that we have a noise threshold below which preparation contextuality holds, we use it to establish a condition for which preparation contextuality is present in the case where the probability of success exceeds that achieved by any classical strategy.   
\end{abstract}

\maketitle


It is important in quantum foundations to have an appropriate definition of what it means for a given quantum feature to resist an explanation within the classical worldview, \textit{i.e.}, an appropriate notion of nonclassicality.
In this respect, one of the leading notions of nonclassicality is preparation contextuality \cite{Spekkens2005}, which refers to the impossibility of a theory to admit of an ontological model that represents operationally equivalent preparations as identical probability distributions on the ontic state space. 
A first challenge to witness such a notion in actual experiments is that the target operational equivalences to test, that are specified \textit{a priori}, are not in general verified. A solution to this issue is to test contextuality not for these target operational equivalences, but to consider the \textit{a posteriori} operational equivalences holding for the noisy preparations that are actually obtained in the experiment. This solution was first adopted by Mazurek et al. in \cite{Mazurek2016}. It was then also used by Pusey in \cite{Pusey2018} to witness preparation contextuality in what is known as the \textit{simplest nontrivial scenario}, which consists of four preparations and two binary-outcome tomographically complete measurements (Fig.~\ref{simplest_scenario}). 

Despite its success, the approach of considering the a posteriori operational equivalences, that we will refer to as Pusey's approach, is not suitable to treat certain information processing tasks defined in the simplest nontrivial scenario and powered by preparation contextuality, such as the two-bit parity-oblivious multiplexing (POM) \cite{Spekkens2009}, in the presence of noise. The reason is that in those tasks the preparations involved, and so the operational equivalences, must be specified in advance (they guarantee that the parity constraint is satisfied), and it is not possible to reanalyze the data in terms of some other sets of preparations. 
For this reason, in this paper we consider an alternative approach to witness nonclassicality in the simplest nontrivial scenario, and we relate it with Pusey's approach.
More precisely, we consider an approach that still refers to the a priori ideal preparations and that is based on a more general notion of classicality named bounded ontological distinctness for preparations ($BOD_P$) \cite{Chaturvedi2020} \footnote{The simplest nontrivial scenario for testing $BOD_P$ would consist of three rather than four preparations, as shown in \cite{Chaturvedi2020}. Nevertheless we keep the already used terminology of Pusey in \cite{Pusey2018} that is of common use when referring to the scenario under consideration here.}. 


Bounded ontological distinctness for preparations generalizes the notion of preparation noncontextuality, that demands that operational equivalences are mapped to ontological equivalences, insofar as it demands that also \textit{differences} are preserved between the operational theory and the ontological model of the theory. 
Unlike \cite{Chaturvedi2020}, we here define differences in terms of distances: the operational distance is the maximal gap over all measurements between the outcome probabilities of the preparations, and the ontological distance is the total variational distance between the ontic distributions \cite{Marvian2020}.
With this notion of classicality, we can now treat the simplest nontrivial scenario in the presence of noise and still refer to the a priori ideal preparations, 
 by addressing the distance the noisy ones have with respect to them. 

We test $BOD_P$ for the distance between the even and odd parity mixtures of the preparations (\textit{i.e.}, $P_+$ and $P_-$ in Fig.~\ref{simplest_scenario}) in the simplest nontrivial scenario. We take this to be the relevant distance because it quantifies the parity communicated in POM in the presence of noise.  We refer to the case with zero difference between the operational and ontological parity distances as satisfying \textit{parity preservation}.


In order to connect our approach based on parity preservation to Pusey's approach based on preparation noncontextuality, we use a third approach developed by Marvian in \cite{Marvian2020}. Marvian's approach quantifies preparation contextuality in terms of the ``inaccessible information'' of an ontological model,
defined as the largest distance between a pair of ontic distributions associated to operationally equivalent preparations. 

In summary, in this paper we present several results on the connection between these three approaches and on the application of such connection: 

\begin{enumerate}
	\item Focusing on the simplest nontrivial scenario, we compare the noise thresholds for witnessing preparation contextuality in Marvian's and Pusey's approaches. We find that Marvian's is more robust\footnote{These thresholds depend on how the noise is modelled (see discussion on this in Section \ref{Conclusion}). The better robustness of Marvian's inequality is due to the better precision in bounding the relevant parameters in terms of the noise $\delta$.}, in that it detects contextuality given $\delta<0.1$, while Pusey's requires $\delta<0.06$ (Theorems \ref{marvian_theorem} and \ref{pusey_theorem}). Here, the noise parameter $\delta$ represents the maximum allowed deviation of the noisy experimental preparations from the corresponding noiseless target preparations in terms of maximum distinguishability with a one-shot measurement. This means that, in a one-shot measurement, the noisy preparations cannot be distinguished from the ideal preparations with probability greater than $\frac{1+\delta}{2}$. 
	
	\item We relate preparation contextuality as witnessed in Marvian's approach and violations of parity preservation. More precisely, we provide bounds defined in terms of the operational statistics for which a violation of Marvian's equality implies a violation of parity preservation and vice versa (Theorem~\ref{relationship_theorem}). 
	
	\item We rewrite the above result in terms of a noise bound by taking into account the noise parameter $\delta$ (Theorem~\ref{relationship_theorem_noise}).  
	As a consequence, we also find that under the threshold $\delta=0.007$, the three approaches -- Pusey's, Marvian's and ours -- agree on detecting nonclassicality; therefore an experimenter can choose the approach that they prefer in order to test nonclassicality (Theorem~\ref{threshold_all}). 
	
	\item 
	We apply the results above to the case of POM in the presence of noise, which is played in the simplest nontrivial scenario. It has been shown that winning the game with a probability of success greater than $\frac{3}{4} + \frac{\varepsilon}{4}$ (which can be achieved with a strategy involving classical bits) implies a violation of $BOD_P$ \cite{Chaturvedi2020}, and more precisely, of parity preservation (Theorem~\ref{theorem_POM}). Here $\varepsilon$ corresponds to the parity communicated as a consequence of the noise. Given that the threshold for violating the parity preservation is $\delta<0.007$, and that below this threshold preparation noncontextuality is also violated (Theorem~\ref{threshold_all}), we have a condition for which POM in the presence of noise is powered by preparation contextuality, thus extending the result of the noiseless case \cite{Spekkens2009}.
	This finding is not at all obvious, given that a violation of $BOD_P$ does not, in general, imply a violation of preparation noncontextuality, and a probability of success greater than $\frac{3}{4} + \frac{\varepsilon}{4}$ has been proven only to imply a violation of $BOD_P$, but not of preparation noncontextuality.
	Moreover, we note that in order to obtain such a result, it is insufficient to combine the no-go theorem of the noiseless case (\textit{i.e.}, that a probability of success greater than $\frac{3}{4}$ implies preparation contextuality) with the fact that preparation contextuality holds whenever $\delta\le0.1$. This is because in the noisy POM one can in principle exploit the noise to communicate some parity also in classical strategies, thus improving the probability of success from $\frac{3}{4} $ to $\frac{3}{4} + \frac{\varepsilon}{4}$.
	
\end{enumerate}

In the remainder of the paper, we first describe, in section \ref{SimplestNontrivialScenario}, the simplest nontrivial scenario. In section \ref{OntologicalModels}, we provide the basic background and terminology regarding operational theories, ontological models, preparation noncontextuality and distances. In section \ref{Approaches}, we describe Pusey's, Marvian's and our approach based on $BOD_P$. In section \ref{Results}, we connect these approaches and state our results. In section \ref{EpsilonPom}, we treat the two-bit parity-oblivious multiplexing in the presence of noise, extending the result of the noiseless case. We conclude, in section \ref{Conclusion}, by summarizing our findings and outlining future research.
 
 \section{The simplest nontrivial scenario}
 \label{SimplestNontrivialScenario}
	
\begin{figure}[htbp]
	\centering
		\subfloat[\label{simplest_scenario_a}]
	{\includegraphics[width=.5\textwidth,height=.35\textheight]{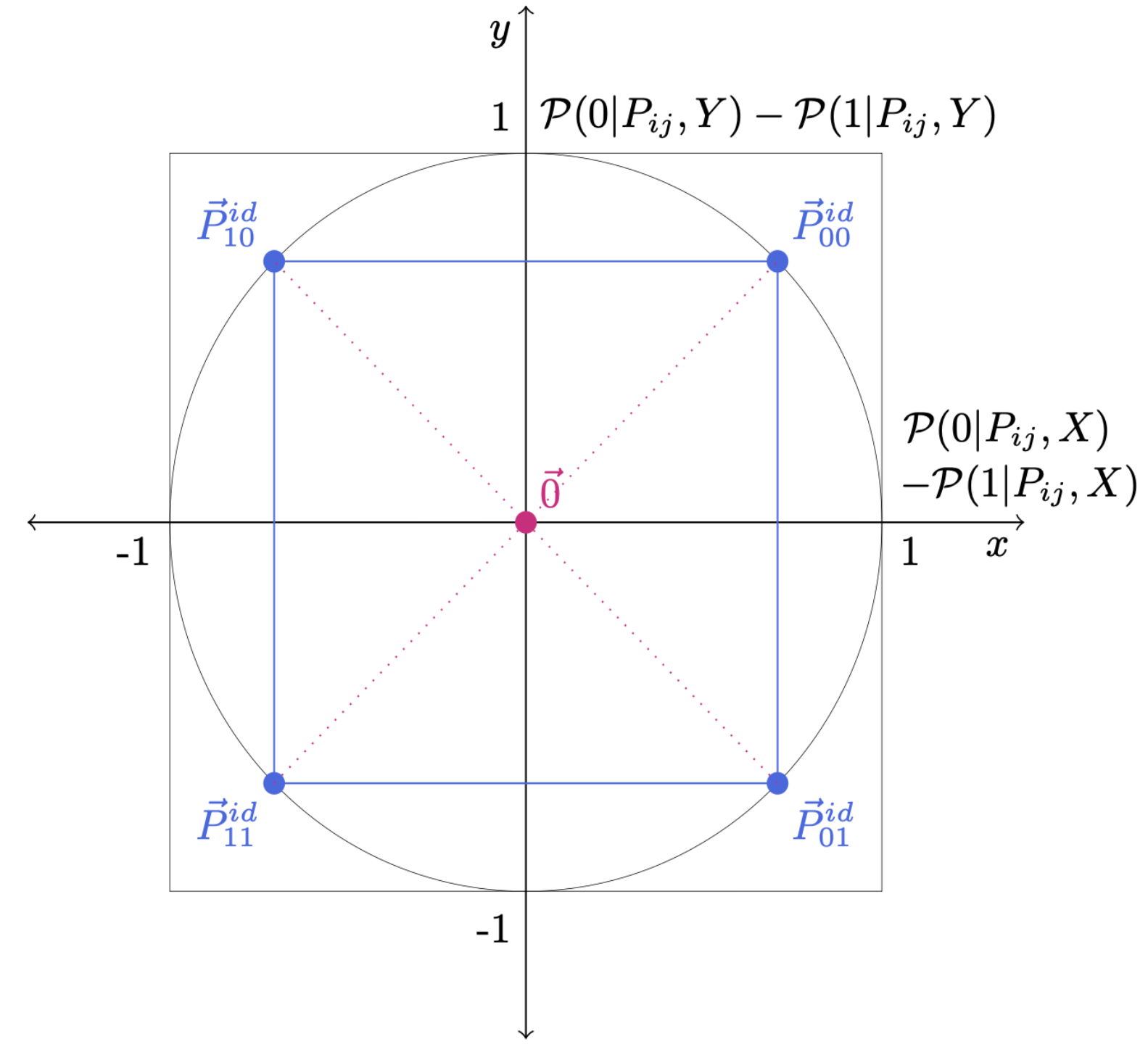}}
		\subfloat[\label{simplest_scenario_b}] 
	{\includegraphics[width=.5\textwidth,height=.35\textheight]{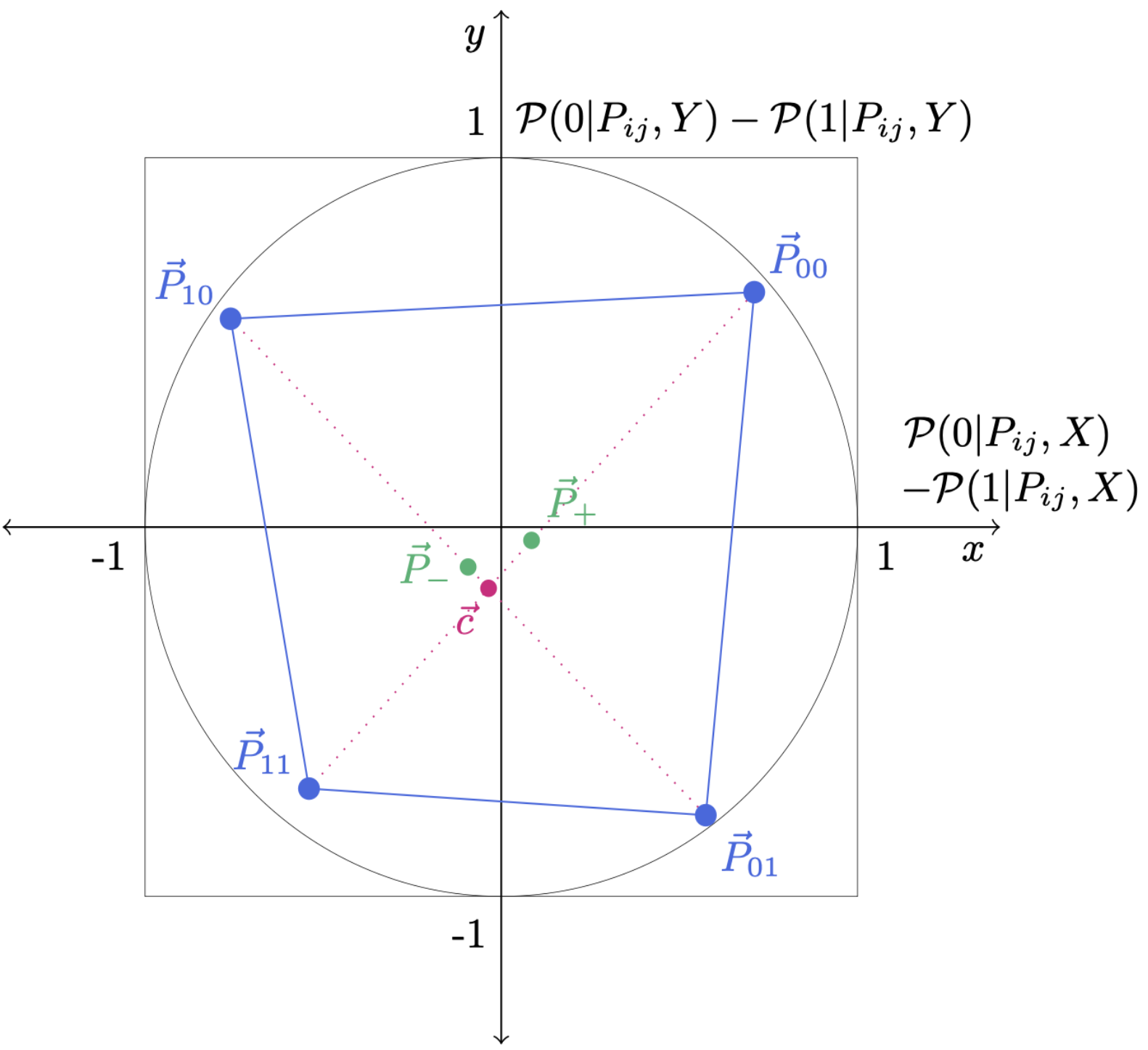}}
		\caption{\textbf{The simplest nontrivial scenario.} Four preparations (vertices of the blue square) and two tomographically complete measurements (corresponding to the $x$ and $y$ axes) are represented within the Bloch circle and the gbit square (in black). 
		Fig.~(a) represents the noiseless case. The a priori operational equivalence of Eq.~\eqref{a priori_O_E} is represented by the vector $\vec{0}$ (in red). 
		Fig.~(b) represents the noisy case. The a posteriori operational equivalence of Eq.~\eqref{Induced_OE} is represented by the vector $\vec{c}$ (in red); the midpoints $\vec{P_+}, \vec{P_-}$ (in green) represent even and odd parity mixtures, respectively.}
	\label{simplest_scenario}
\end{figure}

In this work we focus on what is known as the simplest nontrivial scenario~\cite{Pusey2018}, meaning that we consider theories associated with experiments consisting of four preparations $\{P_{ij}\}=\{P_{00},P_{01},P_{10},P_{11}\}$ and two binary-outcome tomographically complete measurements $\{X,Y\}$.\footnote{The assumption of tomographic completeness guarantees that there cannot exist other measurements that distinguish the operational equivalences between preparations. We refer the reader to \cite{Pusey2019,Mazurek2016,Mazurek2021,Schmid2023addressing} for discussions on how to experimentally deal with the assumption of tomographic completeness.} \footnote{Let us emphasize that $X,Y$ are general measurements and do not represent the Pauli $X,Y$ measurements unless we are explicitly in the case of qubit quantum theory. We adopt this notation for compatibility with the operational statistics being expressed via Cartesian coordinates.} 
The reason why this is the \textit{simplest} nontrivial scenario is that, as shown in appendix B of \cite{Pusey2018}, any other scenario with fewer preparations or measurements always admits the existence of a preparation noncontextual model. A geometrical representation of the scenario is depicted in Fig.~\ref{simplest_scenario}, where the preparations are represented as vectors in the Cartesian plane with the $x$-axis specifying the difference between the probabilities of obtaining outcomes $0$ and $1$ for that preparation given the measurement $X$, and the $y$-axis specifying the same expression but given the measurement $Y$, \textit{i.e.}, for $i,j\in\{0,1\}$,
\begin{align}
\label{x_coordinates}
x_{ij}&=\mathcal{P}(0|P_{ij},X)-\mathcal{P}(1|P_{ij},X),\\
 \label{y_coordinates}
 y_{ij}&=\mathcal{P}(0|P_{ij},Y)-\mathcal{P}(1|P_{ij},Y),\\
 \vec{P}_{ij}&=(x_{ij}, y_{ij}).
\end{align}

\onecolumngrid

\twocolumngrid

Notice that these coordinates take values in $[-1,1]$, and so the preparations can at maximum span the square $\vec{P}_{00}=(1,1),\vec{P}_{01}=(-1,1),\vec{P}_{10}=(1,-1),\vec{P}_{11}=(-1,-1).$ This square corresponds to the gbit state space \cite{barrett2007information,short2010strong}, describing a non-physical theory beyond qubit quantum theory. 

With the case of quantum theory, the four preparations giving the maximum violation of noncontextuality inequalities \cite{Pusey2018,Spekkens2009,Catani2022UR,Catani2023WP} are (modulo the application of a unitary transformation on both preparations and measurements) denoted by $\{P^{id}_{00},P^{id}_{01}, P^{id}_{10}, P^{id}_{11}\}$, and correspond to the vectors
\begin{align}
\label{Ideal_prep}\nonumber
\vec{P}^{id}_{00}\equiv \left(\frac{1}{\sqrt{2}}, \frac{1}{\sqrt{2}}\right), \vec{P}^{id}_{01}\equiv \left(\frac{1}{\sqrt{2}}, -\frac{1}{\sqrt{2}}\right), \\ 
\vec{P}^{id}_{10}\equiv \left(-\frac{1}{\sqrt{2}}, \frac{1}{\sqrt{2}}\right), \vec{P}^{id}_{11}\equiv \left(-\frac{1}{\sqrt{2}}, -\frac{1}{\sqrt{2}}\right).
\end{align}
These are associated to the pure quantum states in the equator of the Bloch sphere located at angles $\frac{\pi}{4}, \frac{3\pi}{4}, \frac{5\pi}{4}, \frac{7\pi}{4}$ with respect to the $+1$ eigenstate of the Pauli $X$ measurement. The superscript \textit{id} stands for ``ideal'', stressing that these are the preparations one aims to prepare in a test of preparation contextuality in the simplest scenario.

Notably, this choice of preparations and measurements also provides the optimal quantum strategy in protocols like the two-bit parity-oblivious multiplexing~\cite{Spekkens2009}, two-bit quantum random access codes~\cite{Ambainis1999}, the CHSH$^*$ game~\cite{CataniHenaut2018} and several others \cite{CataniFaleiro2022}. 

The coordinates (operational statistics) of the vector uniquely determine the preparation it represents. It follows that two preparations $P_a$ and $P_b$ are operationally equivalent -- denoted by $P_a\simeq P_b$ -- if and only if their coordinate vectors are equal: 
\begin{equation}
P_a\simeq P_b \iff \vec{P}_a=\vec{P}_b.
\end{equation}
Notice how the four ideal points $\vec{P}^{id}_{00},\vec{P}^{id}_{01},\vec{P}^{id}_{10}$, and $\vec{P}^{id}_{11}$ form two operationally equivalent decompositions of the preparation represented by the vector $\vec{0}=(0,0)$ at the intersection of the $x$ and $y$ axes (the completely mixed state $\frac{I}{2}$), \textit{i.e.}, 
\begin{equation}\label{a priori_O_E}\frac{1}{2}\vec{P}^{id}_{00}+ \frac{1}{2}\vec{P}^{id}_{11}=\vec{0}=\frac{1}{2}\vec{P}^{id}_{01}+ \frac{1}{2}\vec{P}^{id}_{10}.\end{equation} 
This operational equivalence is termed, the \textit{a priori operational equivalence}, as it is the target  operational equivalence one aims to obtain in an ideal experimental test of preparation contextuality in the simplest scenario prior to performing the actual realistic experiment. 

In realistic experimental scenarios, one cannot exactly prepare the ideal preparations, but necessarily obtains some noisy version of them, which we denote by $\{\vec{P}_{00},\vec{P}_{01},\vec{P}_{10},\vec{P}_{11}\}$.\footnote{Notice that we assume that only the preparations, and not the measurements, are subjected to noise.} 
These obey an operational equivalence different from the one of Eq.~\eqref{a priori_O_E}, that we denote by the vector $\vec{c}$, 
\begin{equation}
\label{Induced_OE}
\underbrace{p\vec{P}_{00}+(1-p)\vec{P}_{11}}_{\vec{P}_p}=\vec{c}=\underbrace{q\vec{P}_{01}+(1-q)\vec{P}_{10}}_{{\vec{P}_q}},
\end{equation} for two probability weights $p,q\in[0,1],$ where we denote the two operationally equivalent preparations as $P_p$ and $P_q$. 
This latter operational equivalence is termed the \textit{a posteriori operational equivalence}.

\section{Ontological models, noncontextuality and distances}
 \label{OntologicalModels}
 
In this section we provide the relevant background material to discuss preparation noncontextuality in the simplest scenario as well as the definitions of the distances that we use. 
\blk
\subsection{Operational theories and ontological models}

The simplest scenario is an example of a prepare and measure scenario. An \textit{operational theory} associated with a prepare and measure scenario is defined by a list of possible preparations, measurements and the probabilities $\mathcal{P}(k|P,M)$ of obtaining the outcome $k$ for the measurement $M$ given that the system is prepared in the preparation $P$. 
An \textit{ontological model} of an operational theory is meant to provide a realist explanation of the operational predictions of the theory~\cite{Harrigan2010}. It does so by stipulating the existence of an \textit{ontic state space} for each given system, denoted with $\Lambda$, which is mathematically represented by a measurable set. Each point $\lambda \in \Lambda$ represents an \textit{ontic state} that describes all the physical properties of the system. 
The ontological model associates each preparation $P$ in the operational theory with a conditional probability distribution $\mu_P(\lambda)\equiv \mu(\lambda|P) $ over ontic states. We call these \textit{epistemic states} as they represent states of knowledge about the underlying ontic states.  Each measurement element $\{k,M\}$ is associated with a conditional probability distribution $\xi(k|\lambda,M)$. The latter corresponds to the probability of obtaining outcome $k$ given that measurement $M$ is implemented on a system in the ontic state $\lambda$. 
An ontological model of an operational theory reproduces the predictions of the theory via the classical law of total probability, 
\begin{align} \label{opdata}
	\mathcal{P}(k|P,M)
	&=\sum_{\lambda\in\Lambda} \xi(k|\lambda,M)\mu(\lambda |P).
\end{align}
Here we consider ontological models for theories associated with the simplest scenario. This means theories whose preparations belong to the convex hull of the four preparations $\{P_{ij}\}$ and whose measurements are the two tomographically complete binary-outcome measurements $X$ and $Y$ that are assumed not to be subjected to noise. 





\subsection{Preparation noncontextuality}

An ontological model is
preparation noncontextual if operationally equivalent preparation procedures are represented by identical probability distributions in the ontological model~\cite{Spekkens2005}. More formally, two preparation procedures $P$ and $P'$ are operationally equivalent if they provide the same operational statistics for all possible measurements, \textit{i.e.}, $\forall M: \mathcal{P}(k|P,M)= \mathcal{P}(k|P',M)$. In this case, we write $P\simeq P'$. An ontological model is preparation noncontextual if any two such preparations are represented by the epistemic states:
\begin{equation}
	P\simeq P' \implies 
	\mu_P= \mu_{P'}.
\end{equation}

An operational theory is termed \textit{preparation noncontextual} if there exists a preparation noncontextual ontological model for the theory.
Notice that, in this paper, we only consider noncontextuality with respect to preparation procedures, but the definition can be applied also to transformation and measurement procedures \cite{Spekkens2005}.



\subsection{Distances}

In order to quantify the distance between two probability distributions 
$\mu_a$ and $\mu_b$ in the ontological model we use the total variational distance, which is the standard choice (see also \cite{Marvian2020}): 
\begin{equation}
d(\mu_a, \mu_b)\equiv\frac{1}{2}\sum_\lambda|\mu_a(\lambda)-\mu_b(\lambda)|.\end{equation}

On the other hand, we define the distances between two preparations $P_a$ and $P_b$ in the operational theory by the largest difference in the probabilities of measurement outcomes between two preparations across all available measurements and respective outcomes in the theory:
\begin{equation}
\label{dist_def_general}
d(P_a, P_b)\equiv \max_{k, M} \{|\mathcal{P}(k|P_a,M)-\mathcal{P}(k|P_b,M)|\}.\\
\end{equation} 

For binary-outcome measurements, this quantity is the same for both the $0$ and $1$ outcome:
\begin{equation}
\begin{aligned}
d(P_a, P_b)&=\max_M \{|\mathcal{P}(0|P_a,M)-\mathcal{P}(0|P_b,M)|\}\\
&=\max_M \{|\mathcal{P}(1|P_a,M)-\mathcal{P}(1|P_b,M)|\}.
\end{aligned}
\end{equation}

Given that there are only two measurements, $X$ and $Y$, the operational distance can be equivalently expressed in the simplest scenario as follows,
\begin{equation}
\label{dist_def_simplest_scenario}
d(P_a, P_b)\equiv\frac{1}{2} \max\{|x_a-x_b|,|y_a-y_b|\}.
\end{equation}

The motivations for employing this specific operational distance are that it provides clear geometrical intuitions and that it makes the calculations tractable, as we will discuss in section \ref{Conclusion}.


\section{Approaches} 
\label{Approaches}
In this section we describe the three approaches considered in this paper: the first is due to Pusey \cite{Pusey2018} and provides robust noncontextuality inequalities for the simplest scenario; the second is due to Marvian \cite{Marvian2020} and witnesses preparation contextuality as the degree of what he defines as inaccessible information; the last is our approach which rephrases bounded ontological distinctness \cite{Chaturvedi2020} in terms of the difference between the operational and ontological distances, and which specifically considers the preparations associated with the even and odd parity mixtures ($P_+$ and $P_-$ in Fig.~\ref{simplest_scenario}), thereby introducing the notion of parity preservation. 


\subsection{Pusey's approach} 

This approach allows one to test preparation contextuality in the simplest scenario considering the a posteriori operational equivalence of Eq.\eqref{Induced_OE} and provides eight noncontextuality inequalities obtained by exploiting a connection to Bell inequalities in the famous Clauser-Horne-Shimony-Holt (CHSH) scenario \cite{CHSH69}.
The requirement of preparation noncontextuality for this a posteriori operational equivalence implies the existence of an ontological model 
for which the epistemic states $\mu_p$ and $\mu_q$ associated with the operationally equivalent preparations $P_p$ and $P_q$ are such that $\mu_p=\mu_q$. 
It is shown in \cite{Pusey2018} that this would be sufficient to make the ontological model preparation noncontextual, \textit{i.e}, that any other operational equivalence within the convex hull of $\{P_{ij}\}$ corresponds to an ontological equivalence in the model. 

Pusey's inequalities  are denoted by $S(x_{ij},y_{ij})\le 0.$ 
In terms of the coordinates $(x_{ij},y_{ij})$ of the noisy preparations, one can write a representative of them as
\begin{multline}
\label{pusey}
S(x_{ij},y_{ij})\\=p(x_{00}+y_{00}+x_{11}+y_{11})+q(x_{01}-y_{01}+x_{10}-y_{10})\\+(y_{10}-x_{10}-x_{11}-y_{11})-2\leq 0.
\end{multline}


This inequality is maximally violated, in quantum theory, by the choice of states and measurements described in the previous section (Fig.~\ref{simplest_scenario_a}), and the expression takes the value $S(x_{ij}^{id},y_{ij}^{id})=4\left(\frac{1}{\sqrt{2}}\right)-2\approx 0.82$. 
The algebraic maximum value of $S(x_{ij},y_{ij})$ is $2$, which is obtained when considering the vertices of the gbit square. 

We emphasize again that Pusey's approach, which considers the a posteriori operational equivalence, is not suitable when considering cases that require the operational equivalence to be specified in advance. An example is the case of the parity-oblivious multiplexing protocol, where the a posteriori operational equivalence does not embody the requirement of parity-obliviousness, unlike the a priori operational equivalence.

\subsection{Marvian's approach}

Marvian's approach quantifies preparation contextuality through the notion of \textit{inaccessible information}, which is defined to be the largest distance between pairs of epistemic states associated to equivalent preparation procedures, minimized over all possible ontological models
\begin{equation}
\label{C_prep}
C_{\prep}^{\min}\equiv \inf_{\Models} \sup_{P_a \simeq P_b} d(\mu_a,\mu_b).
\end{equation}

We note that the inaccessible information $C_{\prep}^{\min}$ of an operational theory is zero if and only if the theory admits of a preparation noncontextual model:
\begin{equation}
\label{marvian_pnc_equality}
C_{\prep}^{\min}=0 \iff \text{ preparation noncontextuality} .
\end{equation}
The equality in Eq.~\eqref{marvian_pnc_equality} is referred to as \textit{Marvian's preparation noncontextuality equality}.

In \cite{Marvian2020}, Marvian provides a lower bound on $C_{\prep}^{\min}$ in terms of operational quantities to witness preparation contextuality. In the simplest scenario, these quantities reduce to a function of the preparations -- here denoted with $\gamma(x_{ij},y_{ij})$ -- the details of which are provided in Appendix \ref{proof2}, 
\begin{equation}
\label{marvian_temp}
C_{\prep}^{\min}\geq \gamma(x_{ij},y_{ij}).
\end{equation}
In contrast to the notion of bounded ontological distinctness and parity preservation that are defined in the next subsections, the approaches of Marvian and Pusey both refer to the same notion: preparation noncontextuality. Therefore, they always agree in detecting contextuality in the noiseless case. However, in the noisy case and given the way we quantify noise, they provide different thresholds below which they are guaranteed to be violated. 
In short, we say that they provide a different \textit{robustification}. This is why they are treated separately and it is important for our purposes to consider both.

 \subsection{$\mathbf{BOD_P}$}
 



$BOD_P$ was introduced in \cite{Chaturvedi2020} and it is a criterion of classicality that requires the equivalence of the operational distinguishability between any two preparations in the operational theory and the ontological distinctness between their ontic representations, generalizing the notion of preparation noncontextuality and being based on the same credentials (a methodological principle inspired by Leibniz \cite{SpekkensLeibniz2019,Schmid2021unscrambling}, or, equivalently, a principle of no operational fine tuning \cite{CataniLeifer2020}). 
The \textit{operational distinguishability} $s_{\mathcal{O}}^{P_a,P_b}$ of a pair of preparations $P_a$ and $P_b$ is defined as 
\begin{equation}
\label{operational_distinguishability}
s_{\mathcal{O}}^{P_a,P_b}\equiv\frac{1}{2}\max_M\left\{\mathcal{P}(0|P_a,M)+\mathcal{P}(1|P_b,M)\right\},
\end{equation}

where the maximum is over all measurements in the operational theory. 

The \textit{ontological distinctness} $s_{\Lambda}^{\mu_a,\mu_b}$ for the corresponding pair of epistemic states $\mu_a$ and $\mu_b$ is defined as
\begin{equation}
\label{ontological_distinctness}
s_{\Lambda}^{\mu_a,\mu_b}\equiv\frac{1}{2} \sum_{\lambda} \max \{\mu_a(\lambda), \mu_b(\lambda)\}.
\end{equation}

We say that an operational theory admits of a model satisfying $BOD_P$ if the value of operational distinguishability for any pair of preparations in the theory equals the value of ontological distinctness for the pair of epistemic states. That is, \textit{bounded ontological distinctness for preparations} demands the following equality to hold for \textit{all pairs} of preparations and corresponding epistemic states:
\begin{equation}
\label{BOD_criteria}
BOD_P \iff s_{\mathcal{O}}^{P_a,P_b}=s_{\Lambda}^{\mu_a,\mu_b} \hspace{0.1cm} \forall a,b.
\end{equation}

To see that a model satisfying $BOD_P$ is preparation noncontextual, we consider any operationally equivalent pair $P_a$ and $P_b$ and observe that the expression in Eq.~\eqref{operational_distinguishability} reduces to $\frac{1}{2}$ when $P_a\simeq P_b$. Given that $s_{\mathcal{O}}^{P_a,P_b}$ represents the maximum probability of distinguishing the two preparations, the value $\frac{1}{2}$ asserts the fact that operationally equivalent procedures are completely indistinguishable. The $BOD_P$ criterion in Eq.~\eqref{BOD_criteria} would then imply that the summation in Eq.~\eqref{ontological_distinctness} equals 1, which entails that $\mu_a(\lambda)=\mu_b(\lambda) \hspace{.2cm} \forall \lambda\in\Lambda$. We therefore have the implication
\begin{equation}
BOD_P \implies \text{preparation noncontextuality}.
\end{equation}

While the above is always true, we emphasize that the converse does not necessarily hold.

In general, an operational theory may not admit of an ontological model satisfying $BOD_P$, thus implying that there exists a pair of preparations for which  $s_{\Lambda}^{\mu_a,\mu_b}-s_{\mathcal{O}}^{P_a,P_b}>0$. The difference $s_{\Lambda}^{\mu_a,\mu_b}-s_{\mathcal{O}}^{P_a,P_b}$ is not explicitly treated by the authors of \cite{Chaturvedi2020}, however it is of crucial relevance in the present paper given that we want to quantify this notion of nonclassicality and consider how it robustifies in the case of realistic noisy scenarios, like the simplest scenario. We quantify the violation of $BOD_P$ as the difference of operational and ontological \textit{distances}. Such quantification can be simply related to the original definition of $BOD_P$ in terms of distinguishability and distinctness, as we now show. Recalling equations \eqref{x_coordinates}, \eqref{y_coordinates}, \eqref{dist_def_simplest_scenario}, and considering the available measurements in the simplest scenario, the relationship between operational distinguishability and distance is established given that 
\begin{equation}
\label{op_dist}
\begin{aligned}
s_{\mathcal{O}}^{P_a,P_b}&=\frac{1}{2}\max_M \left\{\mathcal{P}(0|P_a,M)+\mathcal{P}(1|P_b,M)\right\}\\
&=\max \left\{\frac{1+\frac{1}{2}|x_a-x_b|}{2}, \frac{1+\frac{1}{2}|y_a-y_b|}{2}\right\}\\
&=\frac{1+d(P_a,P_b)}{2}.
\end{aligned}
\end{equation}

A similar relationship holds between the ontological distinctness and distance:
\begin{equation}
\label{ont_dist}
\begin{aligned}
s_{\Lambda}^{\mu_a,\mu_b}&=\frac{1}{2} \sum_{\lambda} \max \left\{\mu_a(\lambda), \mu_b(\lambda)\right\}\\
&=\frac{1}{2}\left(1+\frac{1}{2}\sum_\lambda |\mu_a(\lambda)-\mu_b(\lambda)|\right)\\
&=\frac{1+d(\mu_a,\mu_b)}{2}.
\end{aligned}
\end{equation}

We denote with $\mathcal{D}_{P_a,P_b}$ the difference between the ontological and operational distances for the pair of preparations $P_a$ and $P_b$ and their associated epistemic states. The expression $\mathcal{D}_{P_a,P_b}^{\min}$ denotes the difference $\mathcal{D}_{P_a,P_b}$ minimized over all possible ontological models. That is, 
\begin{align}
\label{dd_def}
\mathcal{D}_{P_a,P_b}&\equiv d(\mu_a,\mu_b)-d(P_a,P_b),\\
\mathcal{D}_{P_a,P_b}^{\min}&\equiv \inf_{\Models} \mathcal{D}_{P_a,P_b}.
\end{align}

Combining equations \eqref{op_dist}, \eqref{ont_dist} and \eqref{dd_def}, we view the difference of operational distinguishability and ontological distinctness as half the difference in operational and ontological distances:
\begin{equation}
s_{\Lambda}^{\mu_a,\mu_b}-s_{\mathcal{O}}^{P_a,P_b}=\frac{1}{2}\mathcal{D}_{P_a,P_b}.
\end{equation}

Therefore $s_{\Lambda}^{\mu_a,\mu_b}-s_{\mathcal{O}}^{P_a,P_b}=0$ if and only if $\mathcal{D}_{P_a,P_b}=0$ and $BOD_P$ can be equivalent expressed as the difference between operational and ontological distances being zero for all pairs:
\begin{equation}
BOD_P \iff \mathcal{D}_{P_a,P_b}^{\min}=0 \hspace{.2cm} \forall P_a,P_b.
\end{equation}





\subsection{Parity preservation}
In this paper we test $BOD_P$ for the difference between the operational and ontological distance, $\mathcal{D}_{P_+,P_-}$, of the even and odd parity mixtures, $P_+= \frac{P_{00}+P_{11}}{2}$ and $P_-= \frac{P_{01}+P_{10}}{2}$ (see Fig.~\ref{simplest_scenario}). The operational distance $d(P_+,P_-)$ indeed codifies the information about the parity between the bits $i$ and $j$ labeling the four preparations $\{P_{ij}\}$ if one measures the preparation with the measurements $X$ and $Y$. For example, if $d(P_+,P_-)=0$, then by measuring $X$ and $Y$ on $P_+$ and $P_-$ one would always get the same outcome, thus obtaining a probability $\frac{1}{2}$ for distinguishing between them and no information about the parity between the bits $i$ and $j$. If $d(P_+,P_-)\neq0$ then some information about the parity between the bits $i$ and $j$ can be obtained by measuring $X$ and $Y$.  If $d(P_+,P_-)$ is preserved in the ontological model, meaning $\mathcal{D}_{P_+,P_-}=0$, we say that there is \textit{parity preservation}. Clearly, a violation of parity preservation implies a violation of $BOD_P$, but not vice versa. If an operational theory does not admit of a parity preserving ontological model, then $\mathcal{D}_{P_+,P_-}^{\min}>0$. 
The focus on parity preservation is relevant in the context of POM and will allow us to connect a violation of $BOD_P$ with a violation of preparation noncontextuality given certain bounds (Theorems \ref{relationship_theorem} and \ref{relationship_theorem_noise}).

\section{Results}
\label{Results}

\begin{figure}[htbp]
	\centering
	{\includegraphics[width=.48\textwidth,height=.35\textheight]{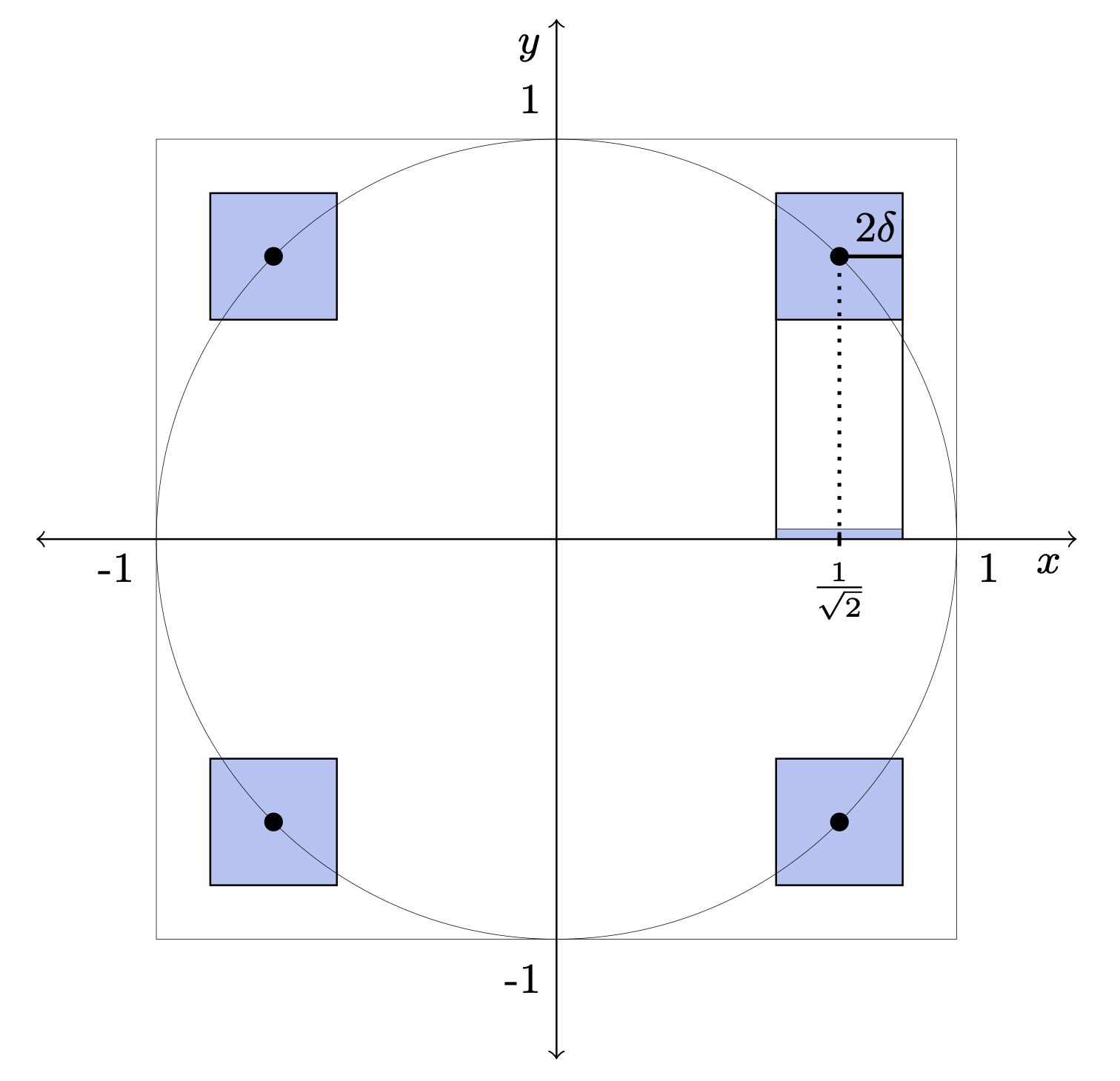}}
	\caption{\textbf{Noise bound of $\delta$}. Ideal points are indicated at the center of the blue squares. These squares of radius $2\delta$ represent regions where noisy points reside, and, in accordance with Eq.~\eqref{dist_def_simplest_scenario}, contain all preparations with an operational distance of \textit{at most} $\delta$ from the ideal points, \textit{i.e.}, $d(\vec{P}_{ij},\vec{P}_{ij}^{id})\leq \delta$. In other words, noisy points in the blue squares cannot be distinguished from their ideal counterparts with a probability greater than $\delta$ in any one-shot measurement.} 
	\label{noise}
\end{figure}

In this section we report the main results (see Table~\ref{results_summary} for a concise summary), all of which are proven in the appendices. 
We begin with Lemma \ref{pusey_lemma}, which reformulates Pusey's preparation noncontextuality inequality with the specification of a noise parameter $\delta$ (see Fig.~\ref{noise}). This enables us to determine, in Theorem \ref{pusey_theorem}, a noise threshold below which the noncontextuality inequality is violated. The same is done for Marvian's preparation noncontextuality equality (Lemma \ref{marvian_lemma}) and a noise threshold is found below which it is still violated (Theorem \ref{marvian_theorem}). We continue by establishing a connection between Marvian's noncontextuality equality and parity preservation in Lemma \ref{lemma_ours_marvian}. As a consequence of this lemma, in Theorem \ref{relationship_theorem}, we provide conditions for which a violation of one notion implies a violation of the other. That is, we show that with enough preparation contextuality, one is certain to violate parity preservation, and vice versa. The parameters for the conditions in Theorem \ref{relationship_theorem} are defined from the experimental data and can be rewritten, with the aid of Lemma \ref{lemma_bounds}, in terms of the noise parameter $\delta$. In this way, Theorem \ref{relationship_theorem} can be rephrased into Theorem \ref{relationship_theorem_noise}. The latter is used to quantify the amount of noise needed to guarantee a violation of parity preservation -- Theorem \ref{threshold_pp}. Finally, having obtained the noise thresholds for violating each notion of nonclassicality considered in this paper, we provide what is arguably the most relevant result -- Theorem \ref{threshold_all} -- that establishes a noise threshold below which all three approaches agree in witnessing nonclassicality.

\subsection{Threshold for violating Pusey's inequality}
\label{section_test}

The results of this subsection are proven in Appendix~\ref{proof1}.

\begin{lemma}
	\label{pusey_lemma}
	Suppose the preparations $\{P_{ij}\}$ of the simplest scenario satisfy a noise bound 
	$d(\vec{P}_{ij},\vec{P}_{ij}^{id})\leq \delta$, where $\delta$ is the noise parameter and $\{P_{ij}^{id}\}$ are the ideal a priori preparations. Pusey's expression $S(x_{ij},y_{ij})$ of Eq.~\eqref{pusey} satisfies the following lower bound in terms of the noise parameter $\delta$: 
	\begin{equation}
	\label{Pusey_Noise_Inequality}
	S(x_{ij},y_{ij})\geq 2\sqrt{2}-2-16\delta+32\sqrt{2}\delta^2,
	\end{equation}
	where $\{x_{ij},y_{ij}\}$ are the coordinates of the preparations $\{P_{ij}\}$.
\end{lemma}

Given that the corresponding preparation noncontextuality inequality is $S(x_{ij},y_{ij})\leq 0$, solving for the right hand side in Eq.~\eqref{Pusey_Noise_Inequality} results in a violation threshold of $\delta \approx0.06$, which leads to the following theorem (see Fig.~\ref{pusey_marvian}).

\begin{theorem}
	\label{pusey_theorem}
	If $d(\vec{P}_{ij},\vec{P}_{ij}^{id})\leq 0.06$, then $S(x_{ij},y_{ij})> 0$ and Pusey's preparation noncontextuality inequality is violated. 
\end{theorem}

\subsection{Threshold for violating Marvian's equality}

The results of this subsection are proven in Appendix~\ref{proof2}.

\begin{lemma}
\label{marvian_lemma}
Suppose the preparations $\{P_{ij}\}$ of the simplest scenario satisfy a noise bound $d(\vec{P}_{ij},\vec{P}_{ij}^{id})\leq \delta$, where $\delta$ is the noise parameter and $\{P_{ij}^{id}\}$ are the ideal a priori preparations. 
Marvian's inaccessible information of Eq.~\eqref{C_prep} of the scenario satisfies the following lower bound in terms of the noise parameter $\delta$:
\begin{equation}
\label{Marvian_Inequality}
C_{\prep}^{\min}\geq \frac{\sqrt{2}-4\delta-1}{4(\sqrt{2}-4\delta)}.
\end{equation}

\end{lemma}

\begin{figure}[htbp]
	\centering
	{\includegraphics[width=.42\textwidth,height=.27\textheight]{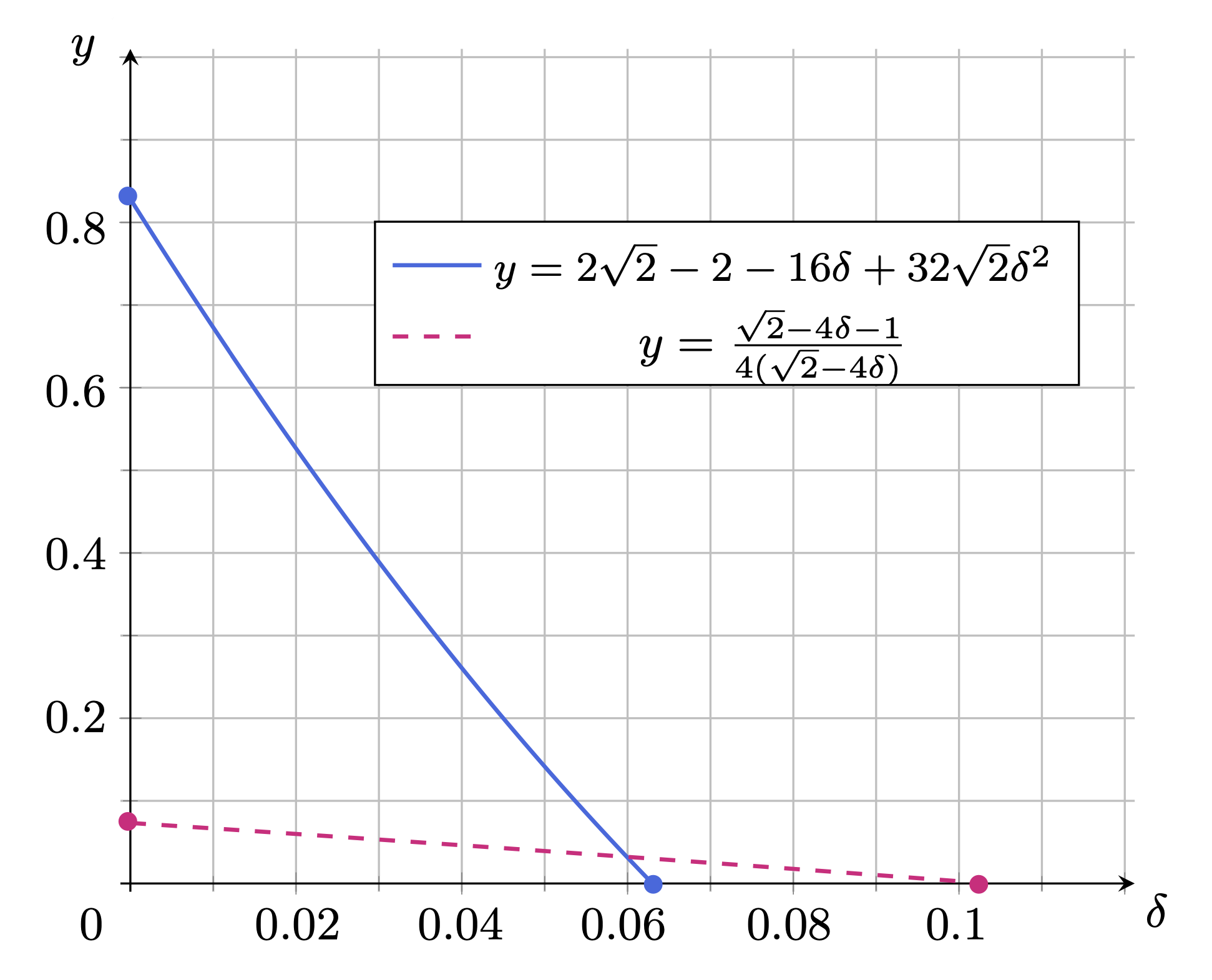}}
	\caption{\textbf{Preparation contextuality witnessed by Pusey's and Marvian's approaches}. 
	The solid blue and dashed red curves plot the functions on the right hand sides of equations \eqref{Pusey_Noise_Inequality} and \eqref{Marvian_Inequality}, respectively. They provide upper bounds to Pusey's and Marvian's expressions, respectively, in terms of the noise parameter $\delta$. 
	What is relevant in the figure is the range of $\delta$ in which each function takes a positive value, thus detecting contextuality. The specific values of the two functions are not to be compared as they are meaningful only within the scope of each approach.
}
\label{pusey_marvian}
\end{figure}

Given that preparation noncontextuality coincides with $C_{\prep}^{\min}=0$, and we obtain a threshold of violation $\delta \approx0.1$ when solving for the right hand side in  Eq.~\eqref{Marvian_Inequality}, we have the following theorem (see Fig.~\ref{pusey_marvian}).

\begin{theorem}
	\label{marvian_theorem}
	If $d(\vec{P}_{ij},\vec{P}_{ij}^{id})\leq 0.1$, then $C_{\prep}^{\min}>0$ and Marvian's preparation noncontextuality equality is violated. 
\end{theorem}


\subsection{Relating Marvian's equality to parity preservation}

We begin by establishing the following inequality, proven in Appendix~\ref{proof3}, connecting $\mathcal{D}_{P_+,P_-}^{\min}$ to $C_{\prep}^{\min}$, thus relating our approach to Marvian's.

\begin{lemma}
\label{lemma_ours_marvian}
	 Given the simplest scenario with even- and odd-parity preparations $P_+$ and $P_-$ 
	 and inaccessible information $C_{\prep}^{\min}$, there exist 
	 functions $\alpha_1,\alpha_2$, and $\alpha_3$ of the preparations $\{P_{ij}\}$ 
	 satisfying
	\begin{subequations}
	\begin{align}
	\label{Distances_Inequality}
	&\mathcal{D}_{P_+,P_-}^{\min}\geq \alpha_1 C_{\prep}^{\min}-\alpha_2, \\
	&\mathcal{D}_{P_+,P_-}^{\min} \leq \alpha_1 C_{\prep}^{\min}+\alpha_3.
	\end{align}
	\end{subequations}
\end{lemma}

By rearranging the terms, these relationships lead to the following theorem. 

\begin{theorem}
	\label{relationship_theorem}
	Given the simplest scenario with even- and odd-parity preparations $P_+$ and $P_-$ 
	and inaccessible information $C_{\prep}^{\min}$, there exist 
	functions $\alpha_1,\alpha_2$, and $\alpha_3$ of the preparations $\{P_{ij}\}$ such that
	\begin{subequations}
		\begin{align}
		\label{implication_1}
		& C_{\prep}^{\min}>\frac{\alpha_2}{\alpha_1} \implies \mathcal{D}_{P_+,P_-}^{\min}>0,\\
		\label{implication_2}
		& \mathcal{D}_{P_+,P_-}^{\min}>\alpha_3 \implies C_{\prep}^{\min}>0.
		\end{align}
	\end{subequations}
	
\end{theorem}


In other words, we have established that a sufficient criterion for violating parity preservation is $C_{\prep}^{\min}>\frac{\alpha_2}{\alpha_1}$ and, in turn, a sufficient criterion for violating Marvian's preparation noncontextuality equality is $\mathcal{D}_{P_+,P_-}^{\min}>\alpha_3$. The parameters $\alpha_1,\alpha_2$, and $\alpha_3$ are evaluated solely from the experimental data and are obtained without making reference to $\delta$. 
Given any four preparations $\{P_{ij}\}$, one can directly calculate the values of $\alpha_1,\alpha_2,$ and $\alpha_3$ and refer to Theorem \ref{relationship_theorem} to see if there is a violation of one notion of classicality given enough violation of the other. Theorem~\ref{relationship_theorem} can be rephrased in terms of the noise parameter $\delta$ via the following lemma, as proven in Appendix~\ref{proof4}.

\begin{lemma}
	\label{lemma_bounds}
	Given the functions $\alpha_1,\alpha_2,$ and $\alpha_3$, if each noisy preparation $P_{ij}$ satisfies $d(\vec{P}_{ij},\vec{P}_{ij}^{id})\leq \delta$, the following upper bounds hold:
	\begin{subequations}
		\begin{align}
		\label{Upper_Bound_a2a1}
		&\frac{\alpha_2}{\alpha_1} \leq \frac{2(1+2\sqrt{3})\delta-4\sqrt{2}\delta^2}{1-2\sqrt{2}\delta},\\
		\label{Upper_Bound_a3}
		&\alpha_3\leq \frac{4\sqrt{3}\delta}{1-2\sqrt{2}\delta-4\sqrt{3}\delta}.
		\end{align}
	\end{subequations}
\end{lemma}

Combining the previous two results, we arrive at the following statement, which recasts Theorem~\ref{relationship_theorem} in terms of the noise bound $\delta$.

\begin{theorem}
		\label{relationship_theorem_noise}
		Given the simplest scenario with even- and odd-parity preparations $P_+$ and $P_-$ 
		and inaccessible information $C_{\prep}^{\min}$, if each $P_{ij}$ satisfies $d(\vec{P}_{ij},\vec{P}_{ij}^{id})\leq \delta$, then the following implications hold:
	\begin{subequations}
		\begin{align}
		\label{implication_1_noise}
		& C_{\prep}^{\min}> \frac{2(1+2\sqrt{3})\delta-4\sqrt{2}\delta^2}{1-2\sqrt{2}\delta} \implies \mathcal{D}_{P_+,P_-}^{\min}>0,\\
		\label{implication_2_noise}
		& \mathcal{D}_{P_+,P_-}^{\min}>\frac{4\sqrt{3}\delta}{1-2\sqrt{2}\delta-4\sqrt{3}\delta} \implies C_{\prep}^{\min}>0.
		\end{align}
	\end{subequations}
\end{theorem}

		


\subsection{Threshold for nonclassicality}

The results in the previous subsection ensure that for any given value of $C_{\prep}^{\min}$, one can find a noise bound $\delta$ such that if noisy preparations $P_{ij}$ lie within $\delta$ distance of the ideal preparations $P_{ij}^{id}$, then $\frac{\alpha_2}{\alpha_1}$ is reduced sufficiently to guarantee that indeed $C_{\prep}^{\min}>\frac{\alpha_2}{\alpha_1}$ and therefore parity preservation is violated. 

\begin{figure}[htbp]
	\centering
	{\includegraphics[width=.45\textwidth,height=.28\textheight]{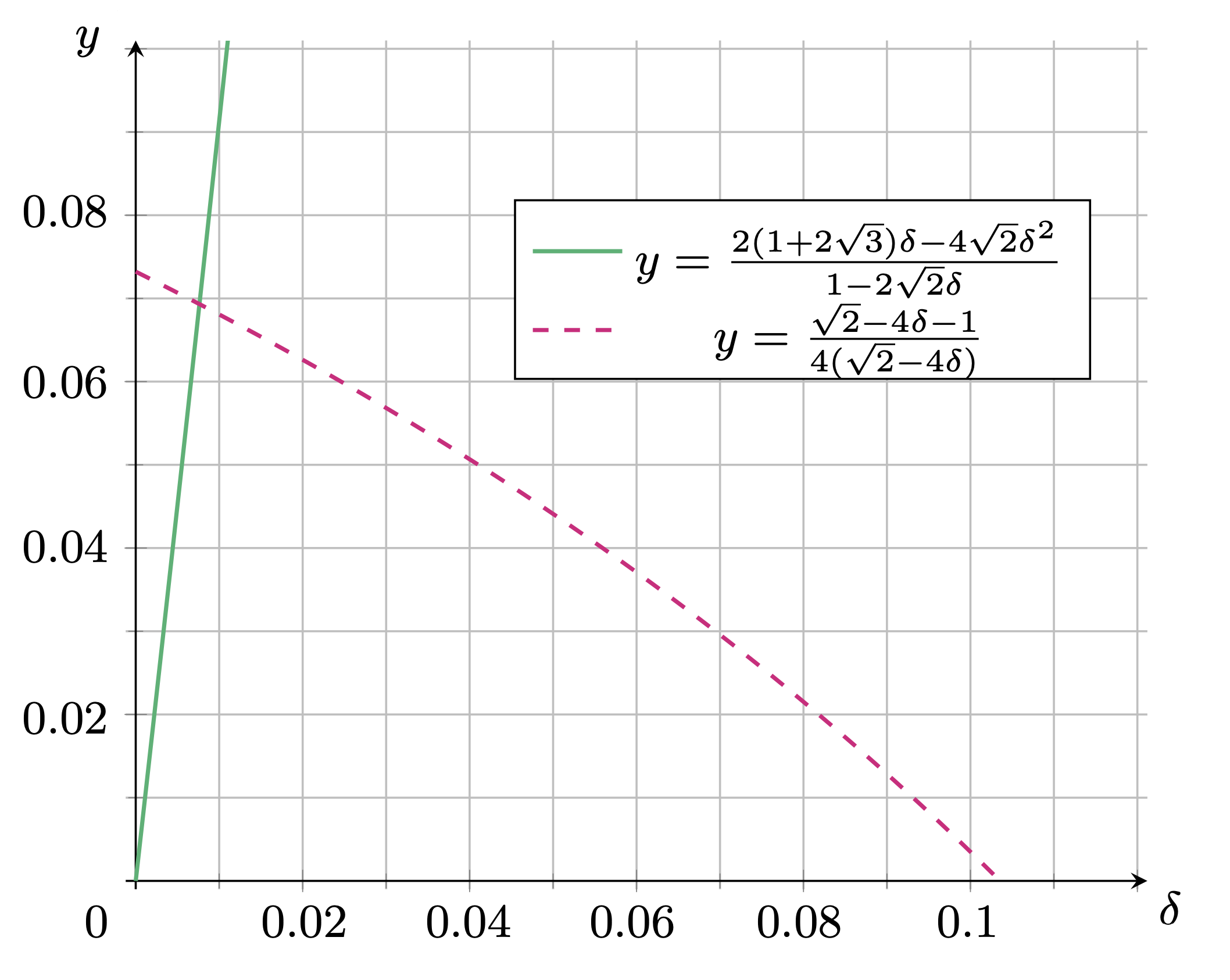}}
	
\caption{\textbf{Violation of parity preservation}. The solid green curve indicates the function of $\delta$ in Eq.~\eqref{implication_1_noise}. For values of $\delta$ smaller than about $0.007$ (see Theorem \ref{threshold_pp}), the dashed red curve upper bounding $C_{\prep}^{\min}$ from Eq.~\eqref{Marvian_Inequality} exceeds the solid green curve and implies, via Eq.~\eqref{implication_1_noise} in Theorem \ref{relationship_theorem_noise}, that parity preservation is violated, \textit{i.e.,} $\mathcal{D}_{P_+,P_-}^{\min}>0$. 
}
\label{parity_preservation_violation}
\end{figure}

Equating the right hand sides of inequalities \eqref{Marvian_Inequality} and \eqref{Upper_Bound_a2a1}, we find the threshold for which Marvian's inequality gives a sufficient lower bound to violate parity preservation as in Eq.~\eqref{implication_1}, which results to be $\delta \approx 0.007$ (see Fig.~\ref{parity_preservation_violation}).

That is, if $\delta \leq 0.007$, then Eq.~\eqref{Upper_Bound_a2a1} gives us $\frac{\alpha_2}{\alpha_1}< 0.063$ whereas Eq.~\eqref{Marvian_Inequality} gives us $C_{\prep}^{\min}> 0.069$, so that indeed $C_{\prep}^{\min}>\frac{\alpha_2}{\alpha_1}$ and $\mathcal{D}_{P_+,P_-}^{\min}$ takes on positive values. We have now established the following theorem.

\begin{theorem}
	\label{threshold_pp}
	If $d(\vec{P}_{ij},\vec{P}_{ij}^{id})\leq 0.007$, then $\mathcal{D}_{P_+,P_-}^{\min}>0$ and parity preservation is violated.
\end{theorem}

Given that Marvian's and Pusey's approaches both exhibit a violation if $\delta\leq0.06$, we can therefore conclude sufficient conditions for equivalency in witnessing nonclassicality between all three approaches as follows.

\begin{theorem}
	\label{threshold_all}
	If the noise parameter $\delta$ satisfies $\delta\leq 0.007$, then $S(x_{ij},y_{ij})>0, C_{\prep}^{\min}>0$, and $\mathcal{D}_{P_+,P_-}^{\min}>0$. Therefore, all three criteria of classicality are violated.
\end{theorem}

In appendix \ref{Appendix_DepolarizingChannel}, we present analogous results for the special case in which the noise manifests as quantum depolarizing noise. In this context, the noise parameter $\delta$ indicates the distance of the noisy preparation from the ideal one along the radial direction (as depicted in Fig.~\ref{noise_depolarization}). For this specific scenario, the noise threshold below which all approaches witness nonclassicality is $0.02.$


\begin{table*}[t!] 
	\centering
	
	\begin{tabular}{|l|l|l|l|l|c|}
		\hline
		Approach      & Notion of nonclassicality &  Reference to a priori ideal preparations & Noise threshold of violation   \\ \hline\hline
		Pusey's          & Preparation contextuality     & No                    & $\delta < 0.06$        \\ \hline
		Marvian's  & Preparation contextuality     & No      & $\delta < 0.1$    \\ \hline   
		This paper & Violation of BOD$_P$  & Yes      & $\delta < 0.007$       \\ \hline   
	\end{tabular}
	\caption{
		\textbf{Three robust ways of witnessing nonclassicality in the simplest scenario.} Violating Pusey's inequality and Marvian's equality are ways of witnessing preparation contextuality. Our approach is based on the notion of parity preservation. A violation of parity preservation is an instance of a violation of $BOD_P$. Violating parity preservation under the indicated threshold means also a violation of preparation noncontextuality.} 
	\label{results_summary}
\end{table*}

\section{Parity-oblivious multiplexing in the presence of noise}
\label{EpsilonPom}

In this section we treat the $m-$bit parity-oblivious multiplexing protocol in both the noiseless and noisy scenarios. In particular, we focus on the $m=2$ case. Indeed, in this instance, the protocol's setting involving four preparations and two measurements corresponds to the simplest scenario of Fig.~\ref{simplest_scenario}.  


\subsection{Noiseless case}

The $m-$bit parity-oblivious multiplexing protocol was first introduced in 2009 by Spekkens \textit{et al} \cite{Spekkens2009}. Since then, it has motivated further investigation and substantial research on relating it to other scenarios \cite{Banik2015,Chailloux2016,Ghorai2018,Saha2019,Ambainis2019,Tavakoli2021,CataniFaleiro2022}, and on developing protocols with preparation contextuality as a resource for the computational advantage \cite{Tavakoli2017,Schmid2018,SahaAnubhav2019,LostaglioSenno2020,lostaglio2020certifying,Yadavalli2020,Flatt2021,Roch2021,Wagner2022}.

As we anticipated, we here consider the case $m=2.$ Let us imagine that Alice prepares a two-bit string, that we denote with $x.$ Bob wishes to learn the value of a single bit among the two (without Alice knowing which one) with a probability at least $p.$ 
Alice and Bob can try to achieve this by agreeing on a strategy that consists of Alice sending some information carriers and Bob performing certain measurements. However, the task contains an additional constraint, called \textit{parity-obliviousness}:  Alice cannot communicate the parity of the two-bit string $x$ to Bob. 
Let us denote the bit that Bob outputs as $b.$ The integer $y$ denotes which of the two bits $b$ should correspond to, and $x_y$ denotes the actual bit in Alice's string.

The probability of success of the game takes, in general, the following form in terms of the probabilities $P(b|P_x,M_y)$, where $P_x$ denotes the preparations of Alice, given the bit string $x$ she has to communicate, and $M_y$ denotes the measurements of Bob, given the bit $y$ to be guessed, 

\begin{equation}
\begin{aligned}
\label{p_suc}
p(b=x_y) &=\frac{1}{8} [\mathcal{P}(0|P_{00},M_0)+\mathcal{P}(0|P_{01},M_0)\\
&\quad+\mathcal{P}(1|P_{10},M_0) +\mathcal{P}(1|P_{11},M_0)\\
&\quad+\mathcal{P}(0|P_{00},M_1)+\mathcal{P}(0|P_{10},M_1)\\
& \quad+\mathcal{P}(1|P_{01},M_1)+\mathcal{P}(1|P_{11},M_1)].
\end{aligned}
\end{equation}

The optimal classical probability of success satisfies $p(b=x_y)\le \frac{3}{4},$ as the only classical encoding that transfers some information to Bob without violating parity-obliviousness consists of encoding only a single bit $x_i.$ Given that $y$ is chosen at random, any bit $x_i$ would perform the same. Therefore, Alice and Bob can agree on Alice always sending $x_1$ and Bob outputting $b=x_1.$ The probability of success is given by the probability that $y=1,$ which is $\frac{1}{2},$ and the probability that Bob outputs correctly (at random, with probability $\frac{1}{2}$) in the other case where $y\neq1,$ that occurs with probability $\frac{1}{2}.$ For this optimal classical strategy we obtain $p(b=x_y)=\frac{1}{2}+ \frac{1}{4}=\frac{3}{4},$ as already stated. This value is the same as the Bell bound of the CHSH game and two-bit quantum random access codes \cite{CHSH69,Ambainis1999}. In \cite{Spekkens2009}, Spekkens \textit{et al} proved the following theorem (here stated only for the case $m=2$).

\begin{theorem}
	\label{Theorem_POM_noiseless}
	The optimal success probability in two-bit parity-oblivious multiplexing of any operational theory that admits of a preparation non-contextual ontological model satisfies $p(b=x_y)\le \frac{3}{4}.$
\end{theorem}

This theorem indicates that preparation contextuality is a necessary resource for performing the two-bit parity-oblivious multiplexing protocol with higher success probability than the one achievable by optimal classical strategies. 
Moreover, it turns out that, by using the same optimal quantum strategy of two-bit quantum random access codes \cite{Ambainis1999} (associated with the preparations $P^{id}_{ij}$ and measurements $X$ and $Y$ of Fig.~\ref{simplest_scenario_a}), the probability of success is $\omega_Q(\textrm{POM})=\cos^2(\frac{\pi}{8})\approx0.85.$  It can be shown that this value is the maximum achievable with quantum strategies \cite{Spekkens2009}. Notice that, when Bob measures on the $X$ or $Y$ basis, he cannot gain any information about the parity, as the parity $0$ and parity $1$ mixtures -- $\frac{1}{2}P^{id}_{00} +\frac{1}{2}P^{id}_{11}$ and  $\frac{1}{2}P^{id}_{01} +\frac{1}{2}P^{id}_{10},$ respectively -- correspond to the same quantum state (the completely mixed state). 

\subsection{Noisy case}

The two-bit parity-oblivious multiplexing protocol in realistic scenarios necessarily involves noisy preparations, that allow some parity to possibly be communicated to Bob.
If one wants to generalize Theorem~\ref{Theorem_POM_noiseless} of the noiseless case and show that preparation contextuality still powers the protocol in the noisy scenario, one faces a couple of challenges. First, one must assume that the noise could potentially be used to communicate parity, thus allowing classical strategies to achieve a probability of success greater than the ones of the noiseless case. Then, once one finds the value of the optimal classical probability of success in the noisy case, one must show that performing better than that implies a proof of preparation contextuality. 

We denote the two-bit parity-oblivious multiplexing protocol in the presence of noise with $\varepsilon-$POM, where $\varepsilon$ denotes the noise in terms of the maximum probability of parity communicated, \textit{i.e.}, the operational distance $d(P_+,P_-)$ between the even- and odd-parity mixtures. 
We first notice that, in order to witness the possible nonclassicality associated with a certain probability of success, Pusey's approach is not ideal. It considers a posteriori operational equivalences, thus not explicitly referring to the parity that can be communicated as a consequence of the noise and which may be part of the reason for the given probability of success. 
We now show how our approach based on the notion of parity preservation, that explicitly refers to the violation of parity obliviousness, is more suitable for the task. 

In order to find an optimal classical strategy in $\varepsilon-$POM, we can mirror the optimal strategy described in the noiseless case of the previous subsection, this time taking into account the $\varepsilon$ parity that can be communicated. Namely, we take into account that Bob knows the parity of $x$, that is, $x_1+ x_2$ (the addition is modulo $2$), with probability $\varepsilon$. 
Without loss of generality, we can assume that Alice and Bob always agree on Alice sending the first bit, $x_1$. If $y=1$, then Bob outputs $b=x_1$ and wins with probability $1$. If $y=2$, then with probability $\varepsilon$ Bob knows the parity and therefore the value of $x_2$ (\textit{i.e.}, he outputs $b=x_1+(x_1+ x_2)=x_2$ and wins with probability 1), and with probability $(1-\varepsilon)$ he does not know the parity and can at best make a random guess about the value of $x_2$ and so win with probability $\frac{1}{2}$. In summary, this strategy results in the probability of success $p(b=x_y)= \frac{1}{2}\times 1+\frac{1}{2}\left[\varepsilon\times 1+(1-\varepsilon) \times \frac{1}{2}\right]= \frac{3}{4}+\frac{\varepsilon}{4}.$ This leads to the following lemma.\footnote{Notice that there is no classical strategy that can perform better than the one we described here. We indeed assumed that Bob uses the knowledge of parity to his maximum possible advantage and that Alice communicates a bit of information, which is the best she can do.}


\begin{manualtheorem}{8}
	The optimal probability of success in two-bit $\varepsilon$-parity-oblivious multiplexing using a classical strategy satisfies $p(b=x_y)\le \frac{3}{4}+\frac{\varepsilon}{4}.$
\end{manualtheorem}

In Appendix \ref{proof6}, we prove the following theorem, which is an application of Proposition 2 and the results of section 4.2 of \cite{Chaturvedi2020}.

\begin{theorem}
	\label{theorem_POM}
	The optimal success probability in two-bit $\varepsilon$-parity-oblivious multiplexing of any operational theory that admits of a parity preserving ontological model satisfies $p(b=x_y)\le \frac{3}{4}+\frac{\varepsilon}{4}.$ 
\end{theorem}

In other words, if one obtains a probability of success greater than the classical probability of success, then parity preservation is violated. Moreover, we have shown in Theorem~\ref{threshold_all} that under a certain threshold, this implies that Marvian's equality (as well as Pusey's inequality) is also violated. Therefore, we have shown that as long as the noise is below a certain threshold, preparation contextuality is still present whenever parity-oblivious multiplexing manifests computational advantage over classical strategies.

\section{Conclusion}
\label{Conclusion}

One essential desideratum for a good notion of nonclassicality is that it should be experimentally testable. Motivated by this, we examined three approaches to test nonclassicality in the simplest nontrivial scenario, which involves four noisy preparations and two tomographically complete measurements. Specifically,  we investigated Pusey's and Marvian's approaches for witnessing preparation contextuality, along with an approach for witnessing a violation of $BOD_P$. 


We showed that these three approaches align in detecting nonclassicality as long as the level of experimental noise remains below a certain threshold, $\delta<0.007$ (in the case of quantum depolarizing noise, this improves to $\delta<0.02$). Therefore, experimenters have the flexibility to choose the approach that best suits their needs when testing for nonclassicality in their experiments, provided the noise remains within this range. This flexibility becomes particularly relevant in scenarios where certain approaches are not suitable, such as in the noisy parity-oblivious multiplexing protocol. In the latter case, we argued that the appropriate notion to test is parity preservation, that refers to the a priori ideal preparations and explicitly allows one to quantify the violation of the parity constraint.  Nevertheless, by virtue of our results, below the noise threshold $\delta<0.007$, Marvian's and Pusey's approaches can also be employed to detect nonclassicality in the experiment. Indeed, below this threshold, we also established that preparation contextuality is still present when one performs the protocol with a success probability greater than what can be achieved with classical strategies.




Crucial to obtain our results is the way we characterized noise through the noise parameter $\delta$. The latter quantifies -- via the operational distance -- the deviation in the measurement statistics between the experimentally realized and the ideal target preparations. Our choice of operational distance corresponds to the maximum difference over the $x$ and $y$ coordinates of the statistics between the preparations. There are two reasons for employing it. First, it is geometrically intuitive, as witnessed by the fact that preparations with $\delta$ noise distance from the ideal ones belong to a square of radius $2\delta$ around those (see Fig.~\ref{noise}). Second, it makes calculations tractable. We give a couple of examples. 

1) Distances over the coordinates are easily identified within Pusey' expression (Eq.~\ref{pusey}), making it possible to bound it. 

2) The operational distance between $P_+$ and $P_-$ (Eq.~\eqref{P+P-_Distance}) coincides with the parity that can be communicated in $\varepsilon-$POM, thus allowing for a straightforward proof of Theorem \ref{theorem_POM}.  

With alternative definitions of operational distance like the maximum relative entropy or the total variational distance \cite{Marvian2020} we would have not exploited the above lucrative features. We leave for future research the question of how the results change by using these other ways of characterizing noise.



It is important to emphasize that determining a mathematical threshold below which both preparation noncontextuality and parity preservation are violated (Theorems \ref{threshold_pp} and \ref{threshold_all}) is not straightforward. While we expected the existence of a noise threshold below which both parity preservation (and consequently $BOD_P$) and preparation noncontextuality are violated, it was not clear how and if this could be found. 
In particular, it was not immediately obvious how to \textit{quantitatively relate} preparation contextuality, that deals with operational equivalences and corresponding ontological inequivalences, with violations of $BOD_P$ (and parity preservation), which deal with operational distances and corresponding greater ontological distances. The key to our achievement in obtaining such a threshold hinges on Lemma \ref{lemma_ours_marvian}. The latter ultimately leads to a function of the noise parameter $\delta$ in Theorem~\ref{relationship_theorem_noise} (Eq.~\eqref{implication_1_noise}) which intersects Marvian's witness of contextuality $C_{\prep}^{\min}$ (as shown in Fig.~\ref{parity_preservation_violation}), revealing a region where not only preparation noncontextuality but also parity preservation is violated. Our success in finding this function lies in the way we recast the operational equivalence in terms of even- and odd-parity mixtures. We managed not only to allow for a connection between preparation noncontextuality and parity preservation but also to retain the amount of parity preservation violation, as shown in Appendix \ref{proof3}. 
Open is the question whether a more precise threshold can be obtained, but we expect this to involve more intricate methods.

Our method based on parity preservation can be seen as an application of the results contained in \cite{Chaturvedi2020} in the context of the simplest scenario. In connection with this previous work, it is important to highlight that we reformulated it using the concept of distances instead of distinguishabilities. This reformulation offers a clear alternative interpretation of both the operational and ontological differences. 
In addition, we reobtained the proof presented in \cite{Chaturvedi2020} about the $\varepsilon-$POM being powered by a violation of $BOD_P$, stressing that the violation is in terms of parity preservation. Consequently, we noticed, by virtue of Theorem \ref{threshold_all}, that $\varepsilon-$POM is also powered by preparation contextuality as long as the noise parameter $\delta$ is below $0.007$. 


The results of this paper are relevant for applications in information processing tasks that aim to witness nonclassicality and that are set in the simplest nontrivial scenario. We recall how these tasks -- examples of which are the two-bit parity-oblivious multiplexing that we treated here and other versions of the two-bit quantum random access codes -- are of central importance because they are the primitive communication tasks where nonclassicality can be certified in a device or semi-device independent way. 
The question remains open as to whether the methods presented in this paper can be extended to scenarios beyond the simplest nontrivial case. We leave this interesting avenue for future research.

\section*{Acknowledgements}

The authors thank Anubhav Chaturvedi for helpful discussions. MK is thankful to David Morrison for helpful discussions. This project started when LC was supported by the Fetzer Franklin Fund of the John E. Fetzer Memorial Trust and by the Army Research Office (ARO) (Grant No. W911NF-18-1-0178). LC also acknowledges funding from the Einstein Research Unit ``Perspectives of a Quantum Digital Transformation" and from the Digital Horizon Europe project FoQaCiA, GA Project No. 101070558. M.L. is grateful for the hospitality of Perimeter Institute where part of this work was carried out. Research at Perimeter Institute is supported in part by the Government of Canada through the Department of Innovation, Science and Economic Development Canada and by the Province of Ontario through the Ministry of Economic Development, Job Creation and Trade.

\bibliography{Bibliography.bib}

\begin{thebibliography}{36}
\expandafter\ifx\csname natexlab\endcsname\relax\def\natexlab#1{#1}\fi
\expandafter\ifx\csname bibnamefont\endcsname\relax
  \def\bibnamefont#1{#1}\fi
\expandafter\ifx\csname bibfnamefont\endcsname\relax
  \def\bibfnamefont#1{#1}\fi
\expandafter\ifx\csname citenamefont\endcsname\relax
  \def\citenamefont#1{#1}\fi
\expandafter\ifx\csname url\endcsname\relax
  \def\url#1{\texttt{#1}}\fi
\expandafter\ifx\csname urlprefix\endcsname\relax\def\urlprefix{URL }\fi
\providecommand{\bibinfo}[2]{#2}
\providecommand{\eprint}[2][]{\url{#2}}

\bibitem[{\citenamefont{Spekkens}(2005)}]{Spekkens2005}
\bibinfo{author}{\bibfnamefont{R.~W.} \bibnamefont{Spekkens}},
  \bibinfo{journal}{Phys. Rev. A} \textbf{\bibinfo{volume}{71}},
  \bibinfo{pages}{052108} (\bibinfo{year}{2005}),
  \urlprefix\url{https://link.aps.org/doi/10.1103/PhysRevA.71.052108}.

\bibitem[{\citenamefont{Mazurek et~al.}(2016)\citenamefont{Mazurek, Pusey,
  Kunjwal, Resch, and Spekkens}}]{Mazurek2016}
\bibinfo{author}{\bibfnamefont{M.~D.} \bibnamefont{Mazurek}},
  \bibinfo{author}{\bibfnamefont{M.~F.} \bibnamefont{Pusey}},
  \bibinfo{author}{\bibfnamefont{R.}~\bibnamefont{Kunjwal}},
  \bibinfo{author}{\bibfnamefont{K.~J.} \bibnamefont{Resch}}, \bibnamefont{and}
  \bibinfo{author}{\bibfnamefont{R.~W.} \bibnamefont{Spekkens}},
  \bibinfo{journal}{Nature Communications} \textbf{\bibinfo{volume}{7}},
  \bibinfo{pages}{ncomms11780} (\bibinfo{year}{2016}),
  \urlprefix\url{https://doi.org/10.1038/ncomms11780}.

\bibitem[{\citenamefont{Pusey}(2018)}]{Pusey2018}
\bibinfo{author}{\bibfnamefont{M.~F.} \bibnamefont{Pusey}},
  \bibinfo{journal}{Phys. Rev. A} \textbf{\bibinfo{volume}{98}},
  \bibinfo{pages}{022112} (\bibinfo{year}{2018}),
  \urlprefix\url{https://link.aps.org/doi/10.1103/PhysRevA.98.022112}.

\bibitem[{\citenamefont{Spekkens et~al.}(2009)\citenamefont{Spekkens, Buzacott,
  Keehn, Toner, and Pryde}}]{Spekkens2009}
\bibinfo{author}{\bibfnamefont{R.~W.} \bibnamefont{Spekkens}},
  \bibinfo{author}{\bibfnamefont{D.~H.} \bibnamefont{Buzacott}},
  \bibinfo{author}{\bibfnamefont{A.~J.} \bibnamefont{Keehn}},
  \bibinfo{author}{\bibfnamefont{B.}~\bibnamefont{Toner}}, \bibnamefont{and}
  \bibinfo{author}{\bibfnamefont{G.~J.} \bibnamefont{Pryde}},
  \bibinfo{journal}{Phys. Rev. Lett.} \textbf{\bibinfo{volume}{102}},
  \bibinfo{pages}{010401} (\bibinfo{year}{2009}),
  \urlprefix\url{https://link.aps.org/doi/10.1103/PhysRevLett.102.010401}.

\bibitem[{\citenamefont{Chaturvedi and Saha}(2020)}]{Chaturvedi2020}
\bibinfo{author}{\bibfnamefont{A.}~\bibnamefont{Chaturvedi}} \bibnamefont{and}
  \bibinfo{author}{\bibfnamefont{D.}~\bibnamefont{Saha}},
  \bibinfo{journal}{{Quantum}} \textbf{\bibinfo{volume}{4}},
  \bibinfo{pages}{345} (\bibinfo{year}{2020}), ISSN \bibinfo{issn}{2521-327X},
  \urlprefix\url{https://doi.org/10.22331/q-2020-10-21-345}.

\bibitem[{\citenamefont{Marvian}(2020)}]{Marvian2020}
\bibinfo{author}{\bibfnamefont{I.}~\bibnamefont{Marvian}},
  \bibinfo{journal}{arXiv:2003.05984v1 [quant-ph]}  (\bibinfo{year}{2020}),
  \urlprefix\url{https://arxiv.org/abs/2003.05984}.

\bibitem[{\citenamefont{Pusey et~al.}(2019)\citenamefont{Pusey, del Rio, and
  Meyer}}]{Pusey2019}
\bibinfo{author}{\bibfnamefont{M.~F.} \bibnamefont{Pusey}},
  \bibinfo{author}{\bibfnamefont{L.}~\bibnamefont{del Rio}}, \bibnamefont{and}
  \bibinfo{author}{\bibfnamefont{B.}~\bibnamefont{Meyer}},
  \bibinfo{journal}{arXiv:1904.08699}  (\bibinfo{year}{2019}),
  \urlprefix\url{https://arxiv.org/abs/1904.08699}.

\bibitem[{\citenamefont{Mazurek et~al.}(2021)\citenamefont{Mazurek, Pusey,
  Resch, and Spekkens}}]{Mazurek2021}
\bibinfo{author}{\bibfnamefont{M.~D.} \bibnamefont{Mazurek}},
  \bibinfo{author}{\bibfnamefont{M.~F.} \bibnamefont{Pusey}},
  \bibinfo{author}{\bibfnamefont{K.~J.} \bibnamefont{Resch}}, \bibnamefont{and}
  \bibinfo{author}{\bibfnamefont{R.~W.} \bibnamefont{Spekkens}},
  \bibinfo{journal}{PRX Quantum} \textbf{\bibinfo{volume}{2}},
  \bibinfo{pages}{020302} (\bibinfo{year}{2021}),
  \urlprefix\url{https://link.aps.org/doi/10.1103/PRXQuantum.2.020302}.

\bibitem[{\citenamefont{Schmid et~al.}(2023)\citenamefont{Schmid, Selby, and
  Spekkens}}]{Schmid2023addressing}
\bibinfo{author}{\bibfnamefont{D.}~\bibnamefont{Schmid}},
  \bibinfo{author}{\bibfnamefont{J.~H.} \bibnamefont{Selby}}, \bibnamefont{and}
  \bibinfo{author}{\bibfnamefont{R.~W.} \bibnamefont{Spekkens}},
  \bibinfo{journal}{arXiv:2302.07282}  (\bibinfo{year}{2023}),
  \urlprefix\url{https://arxiv.org/abs/2302.07282}.

\bibitem[{\citenamefont{Barrett}(2007)}]{barrett2007information}
\bibinfo{author}{\bibfnamefont{J.}~\bibnamefont{Barrett}},
  \bibinfo{journal}{Phys. Rev. A} \textbf{\bibinfo{volume}{75}},
  \bibinfo{pages}{032304} (\bibinfo{year}{2007}),
  \urlprefix\url{https://link.aps.org/doi/10.1103/PhysRevA.75.032304}.

\bibitem[{\citenamefont{Short and Barrett}(2010)}]{short2010strong}
\bibinfo{author}{\bibfnamefont{A.~J.} \bibnamefont{Short}} \bibnamefont{and}
  \bibinfo{author}{\bibfnamefont{J.}~\bibnamefont{Barrett}},
  \bibinfo{journal}{New Journal of Physics} \textbf{\bibinfo{volume}{12}},
  \bibinfo{pages}{033034} (\bibinfo{year}{2010}),
  \urlprefix\url{https://doi.org/10.1088/1367-2630/12/3/033034}.

\bibitem[{\citenamefont{Catani et~al.}(2022)\citenamefont{Catani, Leifer,
  Scala, Schmid, and Spekkens}}]{Catani2022UR}
\bibinfo{author}{\bibfnamefont{L.}~\bibnamefont{Catani}},
  \bibinfo{author}{\bibfnamefont{M.}~\bibnamefont{Leifer}},
  \bibinfo{author}{\bibfnamefont{G.}~\bibnamefont{Scala}},
  \bibinfo{author}{\bibfnamefont{D.}~\bibnamefont{Schmid}}, \bibnamefont{and}
  \bibinfo{author}{\bibfnamefont{R.~W.} \bibnamefont{Spekkens}},
  \bibinfo{journal}{Phys. Rev. Lett.} \textbf{\bibinfo{volume}{129}},
  \bibinfo{pages}{240401} (\bibinfo{year}{2022}),
  \urlprefix\url{https://link.aps.org/doi/10.1103/PhysRevLett.129.240401}.

\bibitem[{\citenamefont{Catani et~al.}(2023)\citenamefont{Catani, Leifer,
  Scala, Schmid, and Spekkens}}]{Catani2023WP}
\bibinfo{author}{\bibfnamefont{L.}~\bibnamefont{Catani}},
  \bibinfo{author}{\bibfnamefont{M.}~\bibnamefont{Leifer}},
  \bibinfo{author}{\bibfnamefont{G.}~\bibnamefont{Scala}},
  \bibinfo{author}{\bibfnamefont{D.}~\bibnamefont{Schmid}}, \bibnamefont{and}
  \bibinfo{author}{\bibfnamefont{R.~W.} \bibnamefont{Spekkens}},
  \bibinfo{journal}{Phys. Rev. A} \textbf{\bibinfo{volume}{108}},
  \bibinfo{pages}{022207} (\bibinfo{year}{2023}),
  \urlprefix\url{https://link.aps.org/doi/10.1103/PhysRevA.108.022207}.

\bibitem[{\citenamefont{Ambainis et~al.}(1999)\citenamefont{Ambainis, Nayak,
  Ta-Shma, and Vazirani}}]{Ambainis1999}
\bibinfo{author}{\bibfnamefont{A.}~\bibnamefont{Ambainis}},
  \bibinfo{author}{\bibfnamefont{A.}~\bibnamefont{Nayak}},
  \bibinfo{author}{\bibfnamefont{A.}~\bibnamefont{Ta-Shma}}, \bibnamefont{and}
  \bibinfo{author}{\bibfnamefont{U.}~\bibnamefont{Vazirani}}, in
  \emph{\bibinfo{booktitle}{Proceedings of the Thirty-First Annual ACM
  Symposium on Theory of Computing}} (\bibinfo{publisher}{Association for
  Computing Machinery}, \bibinfo{address}{New York, NY, USA},
  \bibinfo{year}{1999}), STOC '99, pp. \bibinfo{pages}{376--383}, ISBN
  \bibinfo{isbn}{1581130678},
  \urlprefix\url{https://doi.org/10.1145/301250.301347}.

\bibitem[{\citenamefont{Henaut et~al.}(2018)\citenamefont{Henaut, Catani,
  Browne, Mansfield, and Pappa}}]{CataniHenaut2018}
\bibinfo{author}{\bibfnamefont{L.}~\bibnamefont{Henaut}},
  \bibinfo{author}{\bibfnamefont{L.}~\bibnamefont{Catani}},
  \bibinfo{author}{\bibfnamefont{D.~E.} \bibnamefont{Browne}},
  \bibinfo{author}{\bibfnamefont{S.}~\bibnamefont{Mansfield}},
  \bibnamefont{and} \bibinfo{author}{\bibfnamefont{A.}~\bibnamefont{Pappa}},
  \bibinfo{journal}{Phys. Rev. A} \textbf{\bibinfo{volume}{98}},
  \bibinfo{pages}{060302} (\bibinfo{year}{2018}),
  \urlprefix\url{https://link.aps.org/doi/10.1103/PhysRevA.98.060302}.

\bibitem[{\citenamefont{Catani et~al.}(2024)\citenamefont{Catani, Faleiro,
  Emeriau, Mansfield, and Pappa}}]{CataniFaleiro2022}
\bibinfo{author}{\bibfnamefont{L.}~\bibnamefont{Catani}},
  \bibinfo{author}{\bibfnamefont{R.}~\bibnamefont{Faleiro}},
  \bibinfo{author}{\bibfnamefont{P.-E.} \bibnamefont{Emeriau}},
  \bibinfo{author}{\bibfnamefont{S.}~\bibnamefont{Mansfield}},
  \bibnamefont{and} \bibinfo{author}{\bibfnamefont{A.}~\bibnamefont{Pappa}},
  \bibinfo{journal}{Phys. Rev. A} \textbf{\bibinfo{volume}{109}},
  \bibinfo{pages}{012427} (\bibinfo{year}{2024}),
  \urlprefix\url{https://link.aps.org/doi/10.1103/PhysRevA.109.012427}.

\bibitem[{\citenamefont{Harrigan and Spekkens}(2010)}]{Harrigan2010}
\bibinfo{author}{\bibfnamefont{N.}~\bibnamefont{Harrigan}} \bibnamefont{and}
  \bibinfo{author}{\bibfnamefont{R.~W.} \bibnamefont{Spekkens}},
  \bibinfo{journal}{Foundations of Physics} \textbf{\bibinfo{volume}{40}},
  \bibinfo{pages}{125} (\bibinfo{year}{2010}),
  \urlprefix\url{https://doi.org/10.1007/s10701-009-9347-0}.

\bibitem[{\citenamefont{Clauser et~al.}(1969)\citenamefont{Clauser, Horne,
  Shimony, and Holt}}]{CHSH69}
\bibinfo{author}{\bibfnamefont{J.~F.} \bibnamefont{Clauser}},
  \bibinfo{author}{\bibfnamefont{M.~A.} \bibnamefont{Horne}},
  \bibinfo{author}{\bibfnamefont{A.}~\bibnamefont{Shimony}}, \bibnamefont{and}
  \bibinfo{author}{\bibfnamefont{R.~A.} \bibnamefont{Holt}},
  \bibinfo{journal}{Phys. Rev. Lett.} \textbf{\bibinfo{volume}{23}},
  \bibinfo{pages}{880} (\bibinfo{year}{1969}).

\bibitem[{\citenamefont{Spekkens}(2019)}]{SpekkensLeibniz2019}
\bibinfo{author}{\bibfnamefont{R.~W.} \bibnamefont{Spekkens}},
  \bibinfo{journal}{arXiv:1909.04628}  (\bibinfo{year}{2019}),
  \urlprefix\url{https://arxiv.org/abs/1909.04628}.

\bibitem[{\citenamefont{Schmid et~al.}(2021)\citenamefont{Schmid, Selby, and
  Spekkens}}]{Schmid2021unscrambling}
\bibinfo{author}{\bibfnamefont{D.}~\bibnamefont{Schmid}},
  \bibinfo{author}{\bibfnamefont{J.~H.} \bibnamefont{Selby}}, \bibnamefont{and}
  \bibinfo{author}{\bibfnamefont{R.~W.} \bibnamefont{Spekkens}},
  \bibinfo{journal}{arXiv:2009.03297}  (\bibinfo{year}{2021}),
  \urlprefix\url{https://arxiv.org/abs/2009.03297}.

\bibitem[{\citenamefont{Catani and Leifer}(2023)}]{CataniLeifer2020}
\bibinfo{author}{\bibfnamefont{L.}~\bibnamefont{Catani}} \bibnamefont{and}
  \bibinfo{author}{\bibfnamefont{M.}~\bibnamefont{Leifer}},
  \bibinfo{journal}{{Quantum}} \textbf{\bibinfo{volume}{7}},
  \bibinfo{pages}{948} (\bibinfo{year}{2023}), ISSN \bibinfo{issn}{2521-327X},
  \urlprefix\url{https://doi.org/10.22331/q-2023-03-16-948}.

\bibitem[{\citenamefont{Banik et~al.}(2015)\citenamefont{Banik, Bhattacharya,
  Mukherjee, Roy, Ambainis, and Rai}}]{Banik2015}
\bibinfo{author}{\bibfnamefont{M.}~\bibnamefont{Banik}},
  \bibinfo{author}{\bibfnamefont{S.~S.} \bibnamefont{Bhattacharya}},
  \bibinfo{author}{\bibfnamefont{A.}~\bibnamefont{Mukherjee}},
  \bibinfo{author}{\bibfnamefont{A.}~\bibnamefont{Roy}},
  \bibinfo{author}{\bibfnamefont{A.}~\bibnamefont{Ambainis}}, \bibnamefont{and}
  \bibinfo{author}{\bibfnamefont{A.}~\bibnamefont{Rai}},
  \bibinfo{journal}{Phys. Rev. A} \textbf{\bibinfo{volume}{92}},
  \bibinfo{pages}{030103} (\bibinfo{year}{2015}),
  \urlprefix\url{https://link.aps.org/doi/10.1103/PhysRevA.92.030103}.

\bibitem[{\citenamefont{Chailloux et~al.}(2016)\citenamefont{Chailloux,
  Kerenidis, Kundu, and Sikora}}]{Chailloux2016}
\bibinfo{author}{\bibfnamefont{A.}~\bibnamefont{Chailloux}},
  \bibinfo{author}{\bibfnamefont{I.}~\bibnamefont{Kerenidis}},
  \bibinfo{author}{\bibfnamefont{S.}~\bibnamefont{Kundu}}, \bibnamefont{and}
  \bibinfo{author}{\bibfnamefont{J.}~\bibnamefont{Sikora}},
  \bibinfo{journal}{New Journal of Physics} \textbf{\bibinfo{volume}{18}},
  \bibinfo{pages}{045003} (\bibinfo{year}{2016}),
  \urlprefix\url{https://doi.org/10.1088/1367-2630/18/4/045003}.

\bibitem[{\citenamefont{Ghorai and Pan}(2018)}]{Ghorai2018}
\bibinfo{author}{\bibfnamefont{S.}~\bibnamefont{Ghorai}} \bibnamefont{and}
  \bibinfo{author}{\bibfnamefont{A.~K.} \bibnamefont{Pan}},
  \bibinfo{journal}{Phys. Rev. A} \textbf{\bibinfo{volume}{98}},
  \bibinfo{pages}{032110} (\bibinfo{year}{2018}),
  \urlprefix\url{https://link.aps.org/doi/10.1103/PhysRevA.98.032110}.

\bibitem[{\citenamefont{Saha et~al.}(2019)\citenamefont{Saha, Horodecki, and
  Paw{\l}owski}}]{Saha2019}
\bibinfo{author}{\bibfnamefont{D.}~\bibnamefont{Saha}},
  \bibinfo{author}{\bibfnamefont{P.}~\bibnamefont{Horodecki}},
  \bibnamefont{and}
  \bibinfo{author}{\bibfnamefont{M.}~\bibnamefont{Paw{\l}owski}},
  \bibinfo{journal}{New Journal of Physics} \textbf{\bibinfo{volume}{21}},
  \bibinfo{pages}{093057} (\bibinfo{year}{2019}),
  \urlprefix\url{https://doi.org/10.1088\%2F1367-2630\%2Fab4149}.

\bibitem[{\citenamefont{Ambainis et~al.}(2019)\citenamefont{Ambainis, Banik,
  Chaturvedi, Kravchenko, and Rai}}]{Ambainis2019}
\bibinfo{author}{\bibfnamefont{A.}~\bibnamefont{Ambainis}},
  \bibinfo{author}{\bibfnamefont{M.}~\bibnamefont{Banik}},
  \bibinfo{author}{\bibfnamefont{A.}~\bibnamefont{Chaturvedi}},
  \bibinfo{author}{\bibfnamefont{D.}~\bibnamefont{Kravchenko}},
  \bibnamefont{and} \bibinfo{author}{\bibfnamefont{A.}~\bibnamefont{Rai}},
  \bibinfo{journal}{Quantum Information Processing}
  \textbf{\bibinfo{volume}{18}}, \bibinfo{pages}{111} (\bibinfo{year}{2019}),
  \urlprefix\url{https://doi.org/10.1007/s11128-019-2228-3}.

\bibitem[{\citenamefont{Tavakoli et~al.}(2021)\citenamefont{Tavakoli, Cruzeiro,
  Uola, and Abbott}}]{Tavakoli2021}
\bibinfo{author}{\bibfnamefont{A.}~\bibnamefont{Tavakoli}},
  \bibinfo{author}{\bibfnamefont{E.~Z.} \bibnamefont{Cruzeiro}},
  \bibinfo{author}{\bibfnamefont{R.}~\bibnamefont{Uola}}, \bibnamefont{and}
  \bibinfo{author}{\bibfnamefont{A.~A.} \bibnamefont{Abbott}},
  \bibinfo{journal}{PRX Quantum} \textbf{\bibinfo{volume}{2}},
  \bibinfo{pages}{020334} (\bibinfo{year}{2021}),
  \urlprefix\url{https://link.aps.org/doi/10.1103/PRXQuantum.2.020334}.

\bibitem[{\citenamefont{Hameedi et~al.}(2017)\citenamefont{Hameedi, Tavakoli,
  Marques, and Bourennane}}]{Tavakoli2017}
\bibinfo{author}{\bibfnamefont{A.}~\bibnamefont{Hameedi}},
  \bibinfo{author}{\bibfnamefont{A.}~\bibnamefont{Tavakoli}},
  \bibinfo{author}{\bibfnamefont{B.}~\bibnamefont{Marques}}, \bibnamefont{and}
  \bibinfo{author}{\bibfnamefont{M.}~\bibnamefont{Bourennane}},
  \bibinfo{journal}{Phys. Rev. Lett.} \textbf{\bibinfo{volume}{119}},
  \bibinfo{pages}{220402} (\bibinfo{year}{2017}),
  \urlprefix\url{https://link.aps.org/doi/10.1103/PhysRevLett.119.220402}.

\bibitem[{\citenamefont{Schmid and Spekkens}(2018)}]{Schmid2018}
\bibinfo{author}{\bibfnamefont{D.}~\bibnamefont{Schmid}} \bibnamefont{and}
  \bibinfo{author}{\bibfnamefont{R.~W.} \bibnamefont{Spekkens}},
  \bibinfo{journal}{Phys. Rev. X} \textbf{\bibinfo{volume}{8}},
  \bibinfo{pages}{011015} (\bibinfo{year}{2018}),
  \urlprefix\url{https://link.aps.org/doi/10.1103/PhysRevX.8.011015}.

\bibitem[{\citenamefont{Saha and Chaturvedi}(2019)}]{SahaAnubhav2019}
\bibinfo{author}{\bibfnamefont{D.}~\bibnamefont{Saha}} \bibnamefont{and}
  \bibinfo{author}{\bibfnamefont{A.}~\bibnamefont{Chaturvedi}},
  \bibinfo{journal}{Phys. Rev. A} \textbf{\bibinfo{volume}{100}},
  \bibinfo{pages}{022108} (\bibinfo{year}{2019}),
  \urlprefix\url{https://link.aps.org/doi/10.1103/PhysRevA.100.022108}.

\bibitem[{\citenamefont{Lostaglio and Senno}(2020)}]{LostaglioSenno2020}
\bibinfo{author}{\bibfnamefont{M.}~\bibnamefont{Lostaglio}} \bibnamefont{and}
  \bibinfo{author}{\bibfnamefont{G.}~\bibnamefont{Senno}},
  \bibinfo{journal}{{Quantum}} \textbf{\bibinfo{volume}{4}},
  \bibinfo{pages}{258} (\bibinfo{year}{2020}), ISSN \bibinfo{issn}{2521-327X},
  \urlprefix\url{https://doi.org/10.22331/q-2020-04-27-258}.

\bibitem[{\citenamefont{Lostaglio}(2020)}]{lostaglio2020certifying}
\bibinfo{author}{\bibfnamefont{M.}~\bibnamefont{Lostaglio}},
  \bibinfo{journal}{Phys. Rev. Lett.} \textbf{\bibinfo{volume}{125}},
  \bibinfo{pages}{230603} (\bibinfo{year}{2020}),
  \urlprefix\url{https://link.aps.org/doi/10.1103/PhysRevLett.125.230603}.

\bibitem[{\citenamefont{Yadavalli and Kunjwal}(2022)}]{Yadavalli2020}
\bibinfo{author}{\bibfnamefont{S.~A.} \bibnamefont{Yadavalli}}
  \bibnamefont{and} \bibinfo{author}{\bibfnamefont{R.}~\bibnamefont{Kunjwal}},
  \bibinfo{journal}{{Quantum}} \textbf{\bibinfo{volume}{6}},
  \bibinfo{pages}{839} (\bibinfo{year}{2022}), ISSN \bibinfo{issn}{2521-327X},
  \urlprefix\url{https://doi.org/10.22331/q-2022-10-13-839}.

\bibitem[{\citenamefont{Flatt et~al.}(2022)\citenamefont{Flatt, Lee, Carceller,
  Brask, and Bae}}]{Flatt2021}
\bibinfo{author}{\bibfnamefont{K.}~\bibnamefont{Flatt}},
  \bibinfo{author}{\bibfnamefont{H.}~\bibnamefont{Lee}},
  \bibinfo{author}{\bibfnamefont{C.~R.~I.} \bibnamefont{Carceller}},
  \bibinfo{author}{\bibfnamefont{J.~B.} \bibnamefont{Brask}}, \bibnamefont{and}
  \bibinfo{author}{\bibfnamefont{J.}~\bibnamefont{Bae}}, \bibinfo{journal}{PRX
  Quantum} \textbf{\bibinfo{volume}{3}}, \bibinfo{pages}{030337}
  (\bibinfo{year}{2022}),
  \urlprefix\url{https://link.aps.org/doi/10.1103/PRXQuantum.3.030337}.

\bibitem[{\citenamefont{Roch~i Carceller et~al.}(2022)\citenamefont{Roch~i
  Carceller, Flatt, Lee, Bae, and Brask}}]{Roch2021}
\bibinfo{author}{\bibfnamefont{C.}~\bibnamefont{Roch~i Carceller}},
  \bibinfo{author}{\bibfnamefont{K.}~\bibnamefont{Flatt}},
  \bibinfo{author}{\bibfnamefont{H.}~\bibnamefont{Lee}},
  \bibinfo{author}{\bibfnamefont{J.}~\bibnamefont{Bae}}, \bibnamefont{and}
  \bibinfo{author}{\bibfnamefont{J.~B.} \bibnamefont{Brask}},
  \bibinfo{journal}{Phys. Rev. Lett.} \textbf{\bibinfo{volume}{129}},
  \bibinfo{pages}{050501} (\bibinfo{year}{2022}),
  \urlprefix\url{https://link.aps.org/doi/10.1103/PhysRevLett.129.050501}.

\bibitem[{\citenamefont{Wagner et~al.}(2022)\citenamefont{Wagner, Camillini,
  and Galvao}}]{Wagner2022}
\bibinfo{author}{\bibfnamefont{R.}~\bibnamefont{Wagner}},
  \bibinfo{author}{\bibfnamefont{A.}~\bibnamefont{Camillini}},
  \bibnamefont{and} \bibinfo{author}{\bibfnamefont{E.~F.}
  \bibnamefont{Galvao}}, \bibinfo{journal}{arXiv:2210.05624}
  (\bibinfo{year}{2022}), \urlprefix\url{https://arxiv.org/abs/2210.05624}.

\end{thebibliography}


\appendix

\section{Proof of Lemma \ref{pusey_lemma}}
\label{proof1}

\textit{Suppose the preparations $\{P_{ij}\}$ of the simplest scenario satisfy a noise bound $d(\vec{P}_{ij},\vec{P}_{ij}^{id})\leq \delta$, where $\delta$ is the noise parameter and $\{P_{ij}^{id}\}$ are the ideal a priori preparations. Pusey's expression $S(x_{ij},y_{ij})$ of Eq.~\eqref{pusey} satisfies the following lower bound in terms of the noise parameter $\delta$: 
	\begin{equation}
	S(x_{ij},y_{ij})\geq 2\sqrt{2}-2-16\delta+32\sqrt{2}\delta^2,
	\end{equation}
	where $\{x_{ij},y_{ij}\}$ are the coordinates of the preparations $\{P_{ij}\}$.}

\begin{proof}

Let us consider Pusey's noncontextuality inequality of Eq.~\eqref{pusey} for the case where we insert the coordinates of the preparations $P_{ij}$ satisfying the operational equivalence \eqref{Induced_OE},
\begin{equation}
\begin{aligned}
\label{PuseyIneq}
& p(x_{00}+y_{00}+x_{11}+y_{11})+q(x_{01}-y_{01}+x_{10}-y_{10})\\
& \quad +(y_{10}-x_{10}-x_{11}-y_{11})-2\leq 0.
\end{aligned}
\end{equation}

Suppose $d(\vec{P}_{ij},\vec{P}_{ij}^{id})\leq \delta$. Given the coordinates of $\vec{P}_{ij}^{id}$ specified in Eq.~\eqref{Ideal_prep}, it follows that $l_\delta \leq |x_{ij}|,|y_{ij}|\leq u_\delta$, where $l_\delta =\frac{1}{\sqrt{2}}-2\delta$ and $u_\delta=\frac{1}{\sqrt{2}}+2\delta$. This can be seen in Fig.~\ref{noise}, where any noisy point $\vec{P}_{ij}$ within an operational distance of $\delta$ from the ideal points $\vec{P}_{ij}^{id}$ 
has coordinates whose absolute values lie within the range $\left[\frac{1}{\sqrt{2}}-2\delta, \frac{1}{\sqrt{2}}+2\delta\right]$. Further, $p,q\geq \frac{1-4\sqrt{2}\delta}{2}$ (see Fig.~\ref{lemma_1}).

Applying these bounds to Eq.~\eqref{PuseyIneq}, we obtain

\begin{equation}
\begin{aligned}
S(x_{ij},y_{ij})&=p(x_{00}+y_{00}+x_{11}+y_{11})\\
& \quad +q(x_{01}-y_{01}+x_{10}-y_{10})\\
&\quad \quad +(y_{10}-x_{10}-x_{11}-y_{11})-2\\
&\geq\frac{1-4\sqrt{2}\delta}{2}(l_\delta+l_\delta-u_\delta-u_\delta)\\
& \quad +\frac{1-4\sqrt{2}\delta}{2}(l_\delta+l_\delta-u_\delta-u_\delta)\\
&\quad \quad +(l_\delta+l_\delta+l_\delta+l_\delta)-2\\
&=2(1-4\sqrt{2}\delta)(l_\delta-u_\delta)+4l_\delta-2\\
&=2(1-4\sqrt{2}\delta)\left(\frac{1}{\sqrt{2}}-2\delta-\frac{1}{\sqrt{2}}-2\delta\right)\\
& \quad +4\left(\frac{1}{\sqrt{2}}-2\delta\right)-2\\
&=2\sqrt{2}-2-16\delta+32\sqrt{2}\delta^2.
\end{aligned}
\end{equation}

\begin{figure}[htbp]
	\centering
	{\includegraphics[width=.48\textwidth,height=.35\textheight]{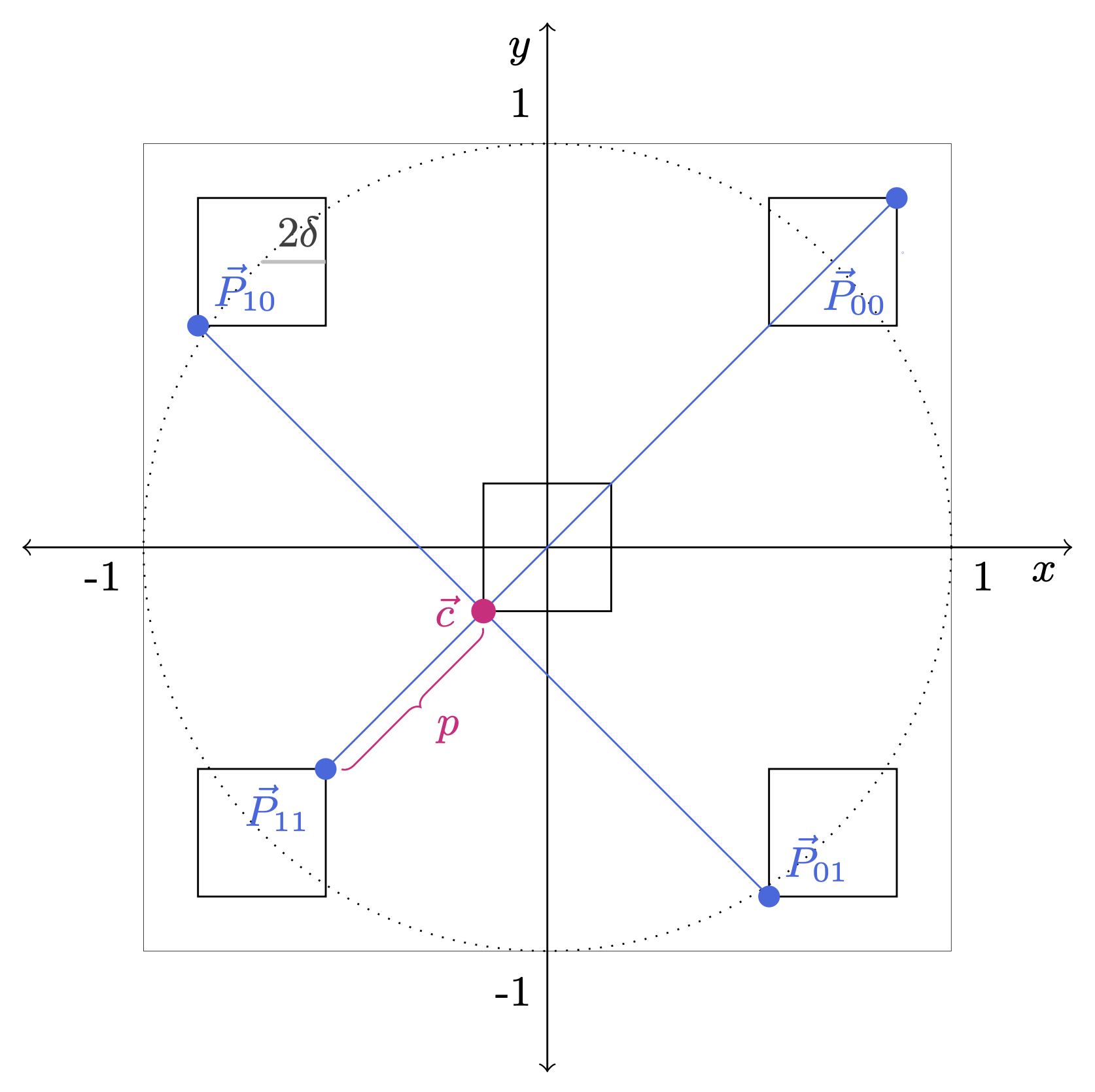}}
	\caption{An example where the weight $p$ in Eq.~\eqref{Induced_OE} takes the minimum possible value, corresponding to $p=\frac{1-4\sqrt{2}\delta}{2}$. This can be seen by noting that $p$ is the weight associated to the distance between $\vec{P}_{11}$ and $\vec{c}$ and observing that the Euclidean distance between $\vec{P}_{11}$ and $\vec{c}$ equals $1-4\sqrt{2}\delta$, while the Euclidean distance between $\vec{P}_{11}$ and $\vec{P}_{00}$ equals 2.}
\label{lemma_1}
\end{figure}

\end{proof}


\section{Proof of Lemma \ref{marvian_lemma}}
\label{proof2}

\textit{Suppose the preparations $\{P_{ij}\}$ of the simplest scenario satisfy a noise bound $d(\vec{P}_{ij},\vec{P}_{ij}^{id})\leq \delta$, where $\delta$ is the noise parameter and $\{P_{ij}^{id}\}$ are the ideal a priori preparations. 
	Marvian's inaccessible information of Eq.~\eqref{C_prep} of the scenario satisfies the following lower bound in terms of the noise parameter $\delta$:
	\begin{equation}
	C_{\prep}^{\min}\geq \frac{\sqrt{2}-4\delta-1}{4(\sqrt{2}-4\delta)}.
	\end{equation}\\}

\begin{figure}[htbp!]
	\centering
	{\includegraphics[width=.46\textwidth,height=.3\textheight]{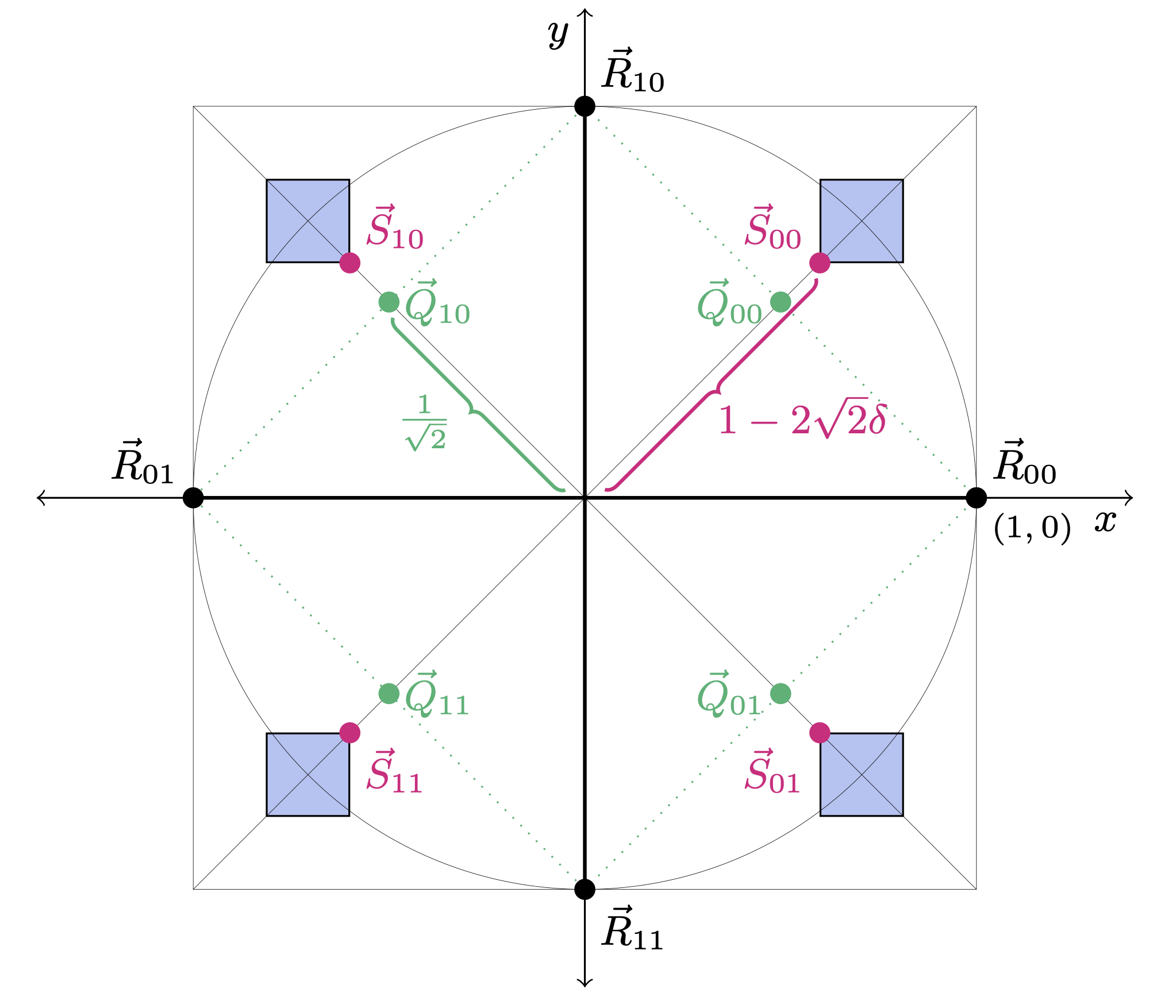}}
\caption{The points $\vec{R}_{ij}$ correspond to the states $R_{ij}$ used to evaluate $P_{\guess}$. Their equal mixtures, denoted $\vec{Q}_{ij}$, are used in calculating $\beta_{\min}$. The set of $\vec{S}_{ij}$ denote the points that are radially furthest from the $\vec{Q}_{ij}$ which are guaranteed to lie within the convex hull of the noisy preparations. That the points $\vec{S}_{ij}$ are radially furthest away from the origin (the completely mixed state) is what ensures $d_{\max}$ to be the largest distance. Note that with less noise (smaller $\delta$), the value of $\sup \{u\}$ increases, since the stationary $\vec{Q}_{ij}$ would then be (relatively) closer to $\vec{0}$ than the $\vec{S}_{ij}$; this in turn decreases $\frac{1}{u}$, which bounds $\beta_{\min}$ from above.}
\label{lemma_2}
\end{figure}

\begin{proof}
	
We begin by using an inequality proven in \cite{Marvian2020}, that provides a lower bound for the inaccessible information of an operational theory:
\begin{equation}
\label{Marvian_a priori_Ineq}
C_{\prep}^{\min}\geq \frac{P_{\guess}-(1-\frac{d-1}{d}\beta_{\min}^{-1})}{(d-1)d^{n-1}}.
\end{equation}

The values $n$ and $d$ represent the number of measurements in the theory and number of outcomes for each measurement, respectively. In the simplest scenario, we have $d=n=2$. The expression $P_{\guess}$ is interpreted in \cite{Marvian2020} as the guessing probability in a certain game therein defined. In the case of the simplest scenario, $P_{\guess}=1$.\footnote {The preparations that give the maximal guessing probability of $1$ are $\vec{R}_{00}=(1,0), \vec{R}_{01}=(-1,0), \vec{R}_{10}=(0,1), \vec{R}_{11}=(0,-1)$. Although these four points are not in our a priori simplest scenario, we can augment the convex hull of the $\{\vec{P}_{ij}\}$s with the set of these additional four points, $U=\{(\pm1,0),(0,\pm1)\}$, as they do not affect or contribute to the contextuality of the scenario. They indeed correspond to the stabilizer states in qubit theory, that are known not to violate preparation noncontextuality inequalities \cite{Catani2022UR}. This can be also seen through Pusey's preparation noncontextuality expression, whereby a) the calculated value of $S(x_{ij},y_{ij})$ for $U$ is zero, and b) the maximal value of contextuality via $S(x_{ij},y_{ij})$ is always attained with the initial four noisy preparations $P_{ij}$. We augment the set of preparations in our simplest scenario with $U$ for the purposes of utilizing Marvian's guessing probability; the inclusion of $U$ allows us to employ the inequality in Eq.~\eqref{Marvian_a priori_Ineq} in a manner that detects nonclassicality appropriately. That is, with $P_{\guess}=1$, the right-hand side of Eq.~\eqref{Marvian_a priori_Ineq} is positive precisely when there is preparation contextuality.}





With these values, we have that Eq.~\eqref{Marvian_a priori_Ineq} reduces to the following:
\begin{equation}
\label{Marvian_a priori_Ineq_ss}
C_{\prep}^{\min}\geq \frac{1}{4}\beta_{\min}^{-1}.
\end{equation}

The remaining term, $\beta_{\min}$, is obtained from the operational statistics and so the right hand side in Eq.~\eqref{Marvian_a priori_Ineq_ss} is just a function of the preparations - it is what we referred to as $\gamma(x_{ij},y_{ij})$ in Eq.~\eqref{marvian_temp}. Let us now define and bound $\beta_{\min}$. Following \cite{Marvian2020}, we have
\begin{equation}
\label{beta_min}
\beta_{\min}\equiv\inf_{P}\max_{i,j} 2^{d_{\max}(P, Q_{ij})},
\end{equation}
where the infimum is taken over all preparations $P$ in the theory, $Q_{ij} \equiv \frac{1}{2} R_{0i}+\frac{1}{2} R_{1j}$ for $i,j \in \{0,1\}$, and the distance $d_{\max}$ is the operational maximum relative entropy for preparations:  
\begin{equation}
\begin{aligned}
&d_{\max}(P_a,P_b)  \equiv-\log_2 \sup \{u:u\leq 1, \exists P_{a'}:\\
& \quad \quad \quad \quad \quad 
 P_b\simeq uP_a+(1-u)P_{a'}\}.
\end{aligned}
\end{equation}

Considering our choices of 
$\vec{R}_{00}=(1,0), \vec{R}_{01}=(-1,0), \vec{R}_{10}=(0,1)$ and $\vec{R}_{11}=(0,-1)$, we have $\vec{Q}_{00}=(\frac{1}{2},\frac{1}{2}), \vec{Q}_{01}=(\frac{1}{2},-\frac{1}{2}), \vec{Q}_{10}=(-\frac{1}{2},\frac{1}{2})$, and $\vec{Q}_{11}=(-\frac{1}{2},-\frac{1}{2})$. We can now bound $\beta_{\min}$ from above:

\begin{equation}
\begin{aligned}
&\beta_{\min}\\
&=\inf_{P}\max_{i,j} 2^{d_{\max}(P, Q_{ij})}\\
&=\inf_{P} \max_{i,j} 2^{-\log_2 \sup\{u:u\leq 1, \exists P_a: Q_{ij}\simeq uP+(1-u)P_a\}}\\
&=\inf_{P} \max_{i,j}(\sup\{u:u\leq 1, \exists P_a:Q_{ij}\simeq uP+(1-u)P_a\})^{-1}\\
&=\max_{i,j}(\sup\{u:u\leq 1, \exists P_a:Q_{ij}\simeq u\frac{I}{2}+(1-u)P_a\})^{-1}\\
&\leq\max_{i,j}(\{u:Q_{ij}\simeq u\frac{I}{2}+(1-u)S_{ij}\})^{-1}\\
&=(\{u:Q_{00}\simeq u\frac{I}{2}+(1-u)S_{00} )^{-1}\\
&= (\{u: \frac{1}{\sqrt{2}}=u\cdot 0+(1-u)(1-2\sqrt{2}\delta)\})^{-1}\\
&=\left(\frac{\sqrt{2}-4\delta-1}{\sqrt{2}-4\delta}\right)^{-1}\\
&=\frac{\sqrt{2}-4\delta}{\sqrt{2}-4\delta-1}.
\end{aligned}
\end{equation}

In the fourth line we use the fact that, as also showed in \cite{Marvian2020}, the infimum is achieved for the completely mixed state $\frac{I}{2}$. The upper bound in the fifth line arises from finding the smallest value that $\sup \{u\}$ can be guaranteed to achieve from mixing $\frac{I}{2}$ with a preparation $P_a$ in our theory to output a fixed $Q_{ij}$. Any $\vec{P}_a$ that lies within the convex hull of the noisy points is a candidate to mix with $\frac{I}{2}$, and the optimal value is achieved with the preparations indicated by $\vec{S}_{ij}$ (see Fig.~\ref{lemma_2}). Due to symmetry, the value of $u$ is the same for any pair of $\vec{Q}_{ij}$ and $\vec{S}_{ij}$. In line 6 we choose, without loss of generality, $ij=00$ to calculate the $u$ value. We conclude with solving $\frac{1}{\sqrt{2}}=(1-u)(1-2\sqrt{2}\delta)$ for $u$.\\

Therefore $\beta_{\min}^{-1}\geq \frac{\sqrt{2}-4\delta-1}{\sqrt{2}-4\delta}$. This inequality applied to Eq.~\eqref{Marvian_a priori_Ineq_ss} establishes the result. 

\end{proof}

\section{Proof of Lemma \ref{lemma_ours_marvian}}
\label{proof3}
	
\textit{Given the simplest scenario with even- and odd-parity preparations $P_+$ and $P_-$ 
	and inaccessible information $C_{\prep}^{\min}$, there exist 
	functions $\alpha_1,\alpha_2,$ and $\alpha_3$ of the preparations $\{P_{ij}\}$ 
satisfying
	\begin{subequations}
		\begin{align}
		&\mathcal{D}_{P_+,P_-}^{\min}\geq \alpha_1 C_{\prep}^{\min}-\alpha_2,\\
		&\mathcal{D}_{P_+,P_-}^{\min} \leq \alpha_1 C_{\prep}^{\min}+\alpha_3.
		\end{align}
	\end{subequations}}

	\begin{proof}
 	We first recall that
 	 \begin{equation*} \underbrace{p\vec{P}_{00}+(1-p)\vec{P}_{11}}_{\vec{P}_p}=\vec{c}=\underbrace{q\vec{P}_{01}+(1-q)\vec{P}_{10}}_{\vec{P}_q},
 	 \end{equation*} and that $\vec{P}_+=\frac{1}{2}\vec{P}_{00}+\frac{1}{2}\vec{P}_{11}$, and  $\vec{P}_-=\frac{1}{2}\vec{P}_{01}+\frac{1}{2}\vec{P}_{10}$.\\
	
	 We then define a weight $r$ and preparations $P_{+'}$ and $P_{-'}$ that pair with $P_+$ and $P_-$ such that the following two criteria hold:
	
(1) $P_{+'}$ and $P_{-'}$ are \textit{also}
		convex combinations of $\{P_{00},P_{11}\}$ and $\{P_{01},P_{10}\}$, respectively.\\

(2) $\vec{P}_p$ and $\vec{P}_q$ can be recast using one common weight $r$ as
		\begin{equation}
		\label{new_combinations} \underbrace{(1-r)\vec{P}_++r\vec{P}_{+'}}_{\vec{P}_p}=\vec{c}=\underbrace{(1-r)\vec{P}_-+r\vec{P}_{-'}}_{\vec{P}_q}.
		\end{equation}
	
	With Eq.~\eqref{new_combinations}, we are able to write $\mu_p=(1-r)\mu_++r\mu_{+'}$ and $\mu_q=(1-r)\mu_-+r\mu_{-'}$. We now show how the quantities just defined allow one to lower and upper bound the distance $d(\mu_p,\mu_q)$, and so $C_{\prep}^{\min}$, in terms of $d(\mu_+,\mu_-)-d(P_+,P_-)$, thus obtaining the wanted result.\\

	We can constructively define the weight $r$ as follows. To start, observe that $\vec{c}$ is either a convex combination of $\{\vec{P}_+, \vec{P}_{00}\}$ or $\{\vec{P}_+, \vec{P}_{11}\}$, and that the weight $p$ determines which combination of the two pairs gives $\vec{c}$. The same is true for $\vec{c}$ being expressed as a convex combination of either $\{\vec{P}_-, \vec{P}_{01}\}$ or $\{\vec{P}_-, \vec{P}_{10}\}$, with the weight $q$ determining which pair. With this in mind, we define the even- and odd-parity weights, $r_+$ and $r_-$, to be (here all magnitudes $||\cdot||$ indicate Euclidean length), 
	\begin{equation}
	r_+\equiv\begin{cases} 
	\frac{||\vec{P}_+ -\vec{c}||}{||\vec{P}_+-\vec{P}_{00}||}& p\geq\frac{1}{2} \\
	\frac{||\vec{P}_+-\vec{c}||}{||\vec{P}_+-\vec{P}_{11}||}& p\leq\frac{1}{2} 
	\end{cases} ,\hspace{.3cm} r_-\equiv\begin{cases} 
	\frac{||\vec{P}_--\vec{c}||}{||\vec{P}_--\vec{P}_{01}||}& q\geq\frac{1}{2} \\
	\frac{||\vec{P}_--\vec{c}||}{||\vec{P}_--\vec{P}_{10}||}& q\leq\frac{1}{2}
	\end{cases}.
	\label{r_values}
	\end{equation}

	We now take $r$ to be the maximum of these two values, as this ensures that our first criterion outlined earlier is satisfied, namely that the preparations $P_{+'}$ and $P_{-'}$ can be expressed as convex combinations of $\{P_{00},P_{11}\}$ and $\{P_{01},P_{10}\}$, respectively. In practice, one of either $P_{+'}$ or $P_{-'}$ will equate to the original noisy $P_{ij}$, with the other corresponding to a strict convex combination (see Fig.~\ref{lemma_3}). The specifics will depend on the weights in Eq.~\eqref{Induced_OE}.
	
	\newpage
	
	\onecolumngrid
	
	\begin{figure}[h!]
		\centering
		{\includegraphics[width=.9\textwidth,height=.7\textheight]{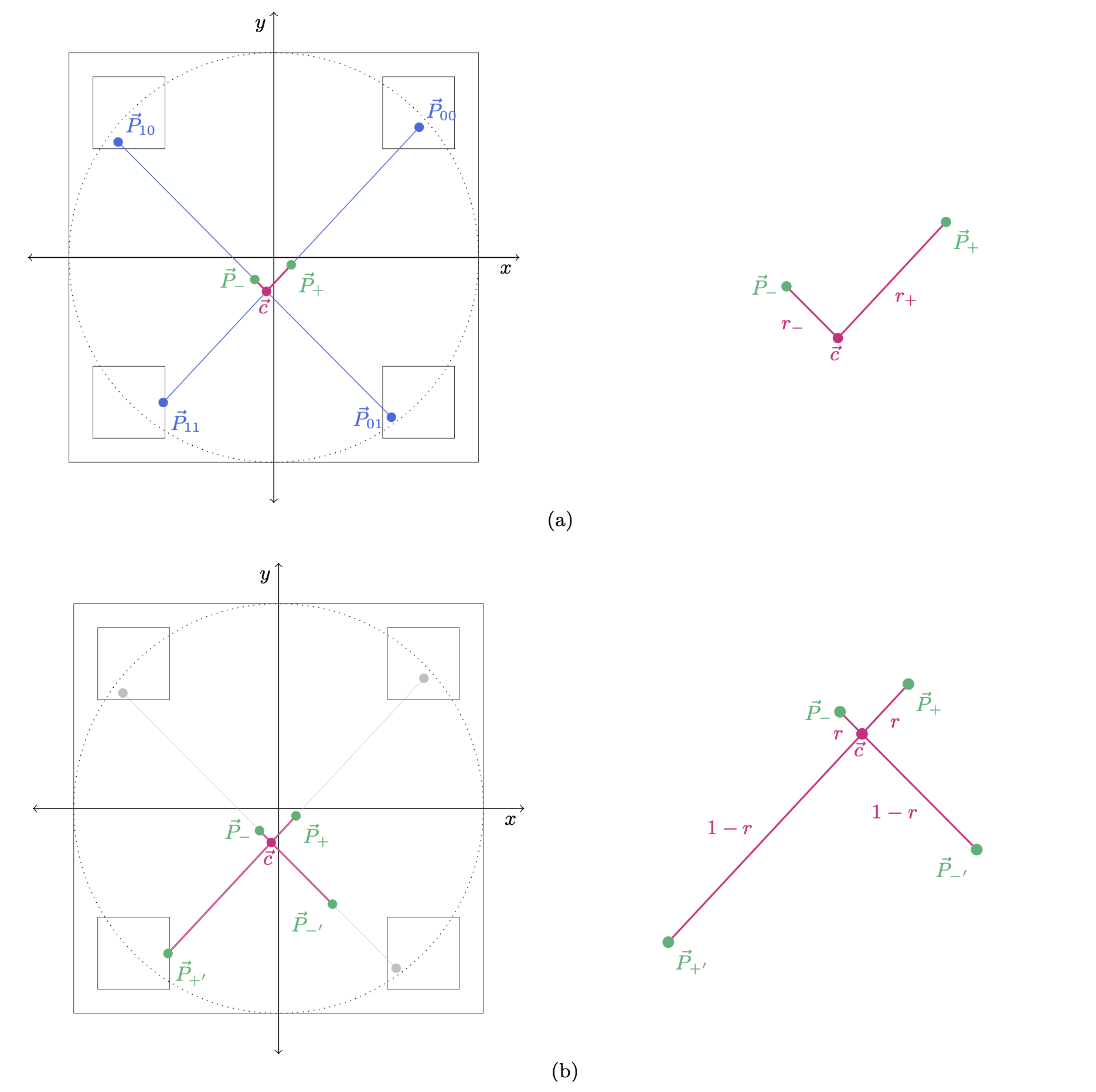}}
		\caption{(a) In this example $p<\frac{1}{2}$ and $q>\frac{1}{2}$, implying that $\vec{c}$ is a combination of $\{\vec{P}_+,\vec{P}_{11}\}$ and $\{\vec{P}_-,\vec{P}_{01}\}$, respectively, as shown on the left side. We also note that $r_+>r_-$, as shown on the right side, so that $r_+$ will be used as the ratio $r$. (b) The intersection $\vec{c}$ has been recast (left side) in terms of two new convex combinations $\{\vec{P}_-,\vec{P}_{-'}\}$ and $\{\vec{P}_+,\vec{P}_{+'}\}$ that use the same weight $r$ (right side). Given that these combinations are themselves mixtures of the a priori preparations in Eq.~\eqref{Induced_OE}, they can be used to express $\vec{P}_p$ and $\vec{P}_q$. We note that in this example, $\vec{P}_{+'}=\vec{P}_{11}$, while $\vec{P}_{-'}$ is now a strictly convex combination of $\vec{P}_-$ and $\vec{P}_{01}$. (Choosing $r$ to have been the minimum rather than maximum would have retained $\vec{P}_{-'}=\vec{P}_{01};$ however, $\vec{P}_{+'}$ would have then been outside of the convex hull of our preparations and thus outside of the operational theory.)}
		\label{lemma_3}
	\end{figure}
	
	\twocolumngrid

	We now have 
	\begin{equation}
	\begin{aligned}
	d(\mu_p,\mu_q)&=d((1-r)\mu_++r\mu_{+'},(1-r)\mu_-+r\mu_{-'})\\
	&\leq (1-r)d(\mu_+,\mu_-)+rd(\mu_{+'},\mu_{-'}) \\
	&\leq (1-r)d(\mu_+,\mu_-)+r ,
	\end{aligned}
	\end{equation}
	
	where the first inequality follows from the triangle inequality and the second inequality follows from  $d(\mu_{+'},\mu_{-'})\leq 1$.\\
	
		It therefore follows that
	\begin{equation}
		\begin{aligned}
			&\frac{1}{1-r}d(\mu_p,\mu_q)\leq d(\mu_+,\mu_-) + \frac{r}{1-r}\\
			&\underbrace{\frac{1}{1-r}}_{\alpha_1} d(\mu_p,\mu_q)-\underbrace{\left[\frac{r}{(1-r)}+d(P_+,P_-)\right]}_{\alpha_2}\\
			& \quad \leq d(\mu_+,\mu_-)-d(P_+,P_-). 
			\label{lower_bound}
		\end{aligned}
	\end{equation}
	
	Similarly, we have
	\begin{equation}
		\begin{aligned}	d(\mu_p,\mu_q)&=d((1-r)\mu_++r\mu_{+'},(1-r)\mu_-+r\mu_{-'})\\
			&\geq (1-r)d(\mu_+,\mu_-)-rd(\mu_{+'},\mu_{-'}) \\
			&\geq (1-r)d(\mu_+,\mu_-)-r,
		\end{aligned}
	\end{equation}
	
		which gives us
	\begin{equation}
		\begin{aligned}
			&\frac{1}{1-r}d(\mu_p,\mu_q)\geq d(\mu_+,\mu_-) - \frac{r}{1-r}\\
			&\underbrace{\frac{1}{1-r}}_{\alpha_1} d(\mu_p,\mu_q)+\underbrace{\left[\frac{r}{(1-r)}-d(P_+,P_-)\right]}_{\alpha_3}\\
			& \quad \geq d(\mu_+,\mu_-)-d(P_+,P_-).
			\label{upper_bound}
		\end{aligned}
	\end{equation}
	
	In the expressions above -- equations \eqref{lower_bound} and \eqref{upper_bound} -- we have identified the functions $\alpha_1,\alpha_2,$ and $\alpha_3.$ Taking the infimum of equations \eqref{lower_bound} and \eqref{upper_bound} over all ontological models, and noting that $C_{\prep}^{\min}$ is the infimum of $d(\mu_p,\mu_q)$ across all models, the desired result is proven.

\end{proof}

\section{Proof of Lemma \ref{lemma_bounds}}
\label{proof4}

\textit{Given the functions $\alpha_1,\alpha_2,$ and $\alpha_3$, if each noisy preparation $P_{ij}$ satisfies $d(\vec{P}_{ij},\vec{P}_{ij}^{id})\leq \delta$, the following upper bounds hold:
	\begin{subequations}
		\begin{align}
		&\frac{\alpha_2}{\alpha_1} \leq \frac{2(1+2\sqrt{3})\delta-4\sqrt{2}\delta^2}{1-2\sqrt{2}\delta}.\\
		&\alpha_3\leq \frac{\frac{4\sqrt{3}\delta}{1-2\sqrt{2}\delta}}{1-\frac{4\sqrt{3}\delta}{1-2\sqrt{2}\delta}}.
		\end{align}
	\end{subequations}}

\begin{figure}[h!]
	\centering
	\subfloat[\label{lemma_4_a}]
	{\includegraphics[width=.5\textwidth,height=.37\textheight]{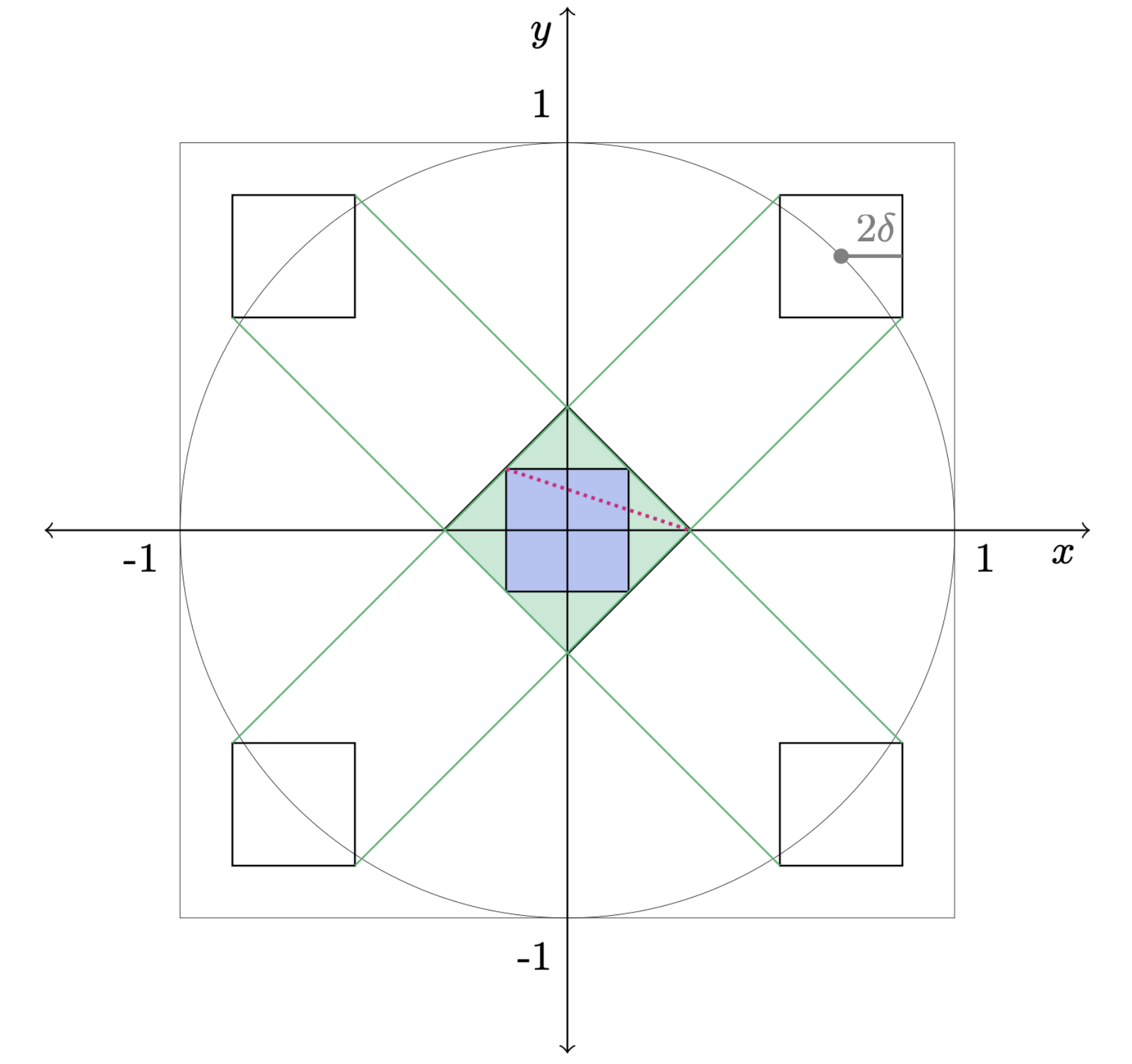}}
	\subfloat[\label{lemma_4_b}]
	{\includegraphics[width=.5\textwidth,height=.37\textheight]{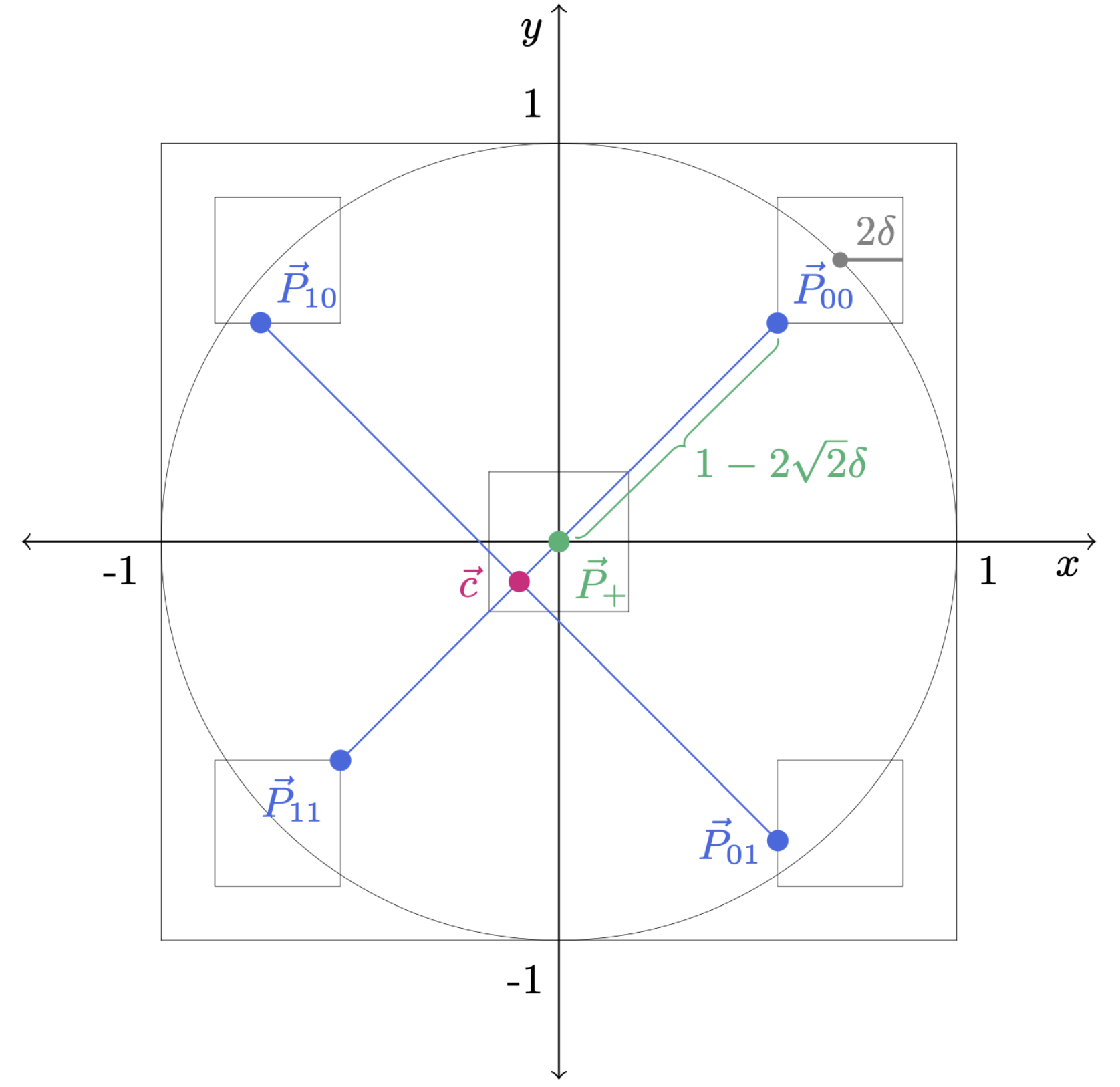}}
	\caption{(a) The point $\vec{c}$ lies within the greater rhombus region shaded in green. The points $\vec{P}_+,\vec{P}_-$ lie within the inscribed square shaded in blue. The dotted red line illustrates the maximal distance between points lying in each region, which is of length $4\sqrt{3}\delta$. (b) An example of the minimum possible Euclidean distance between a preparation vector $\vec{P}_{ij}$ and midpoint vector $\vec{P}_+$ or $\vec{P}_-$, 
		here shown with the case of $\vec{P}_{00}$ and $\vec{P}_+$, where $||\vec{P}_+-\vec{P}_{00}||=1-2\sqrt{2}\delta$.}
	\label{lemma_4}
\end{figure}

\begin{proof}

From Eq.~\eqref{lower_bound}, we have $\alpha_1=\frac{1}{1-r}$ and $\alpha_2=\frac{r}{1-r}+d(P_+,P_-)$, so that

\begin{equation}
\label{a2a1_expression}
\frac{\alpha_2}{\alpha_1}=d(P_+,P_-)(1-r)+r.
\end{equation}

Note that $\vec{P}_+$ and $\vec{P}_-$ are averages of points contained in $\delta$-neighborhoods of the ideal points. Therefore, $\vec{P}_+$ and $\vec{P}_-$ are each contained in a $\delta$-neighborhood of the origin, which implies that
\begin{equation}
d(P_+,P_-)\leq 2\delta.
\end{equation}

Next, we observe that $\vec{c}$ is the intersection point of two line segments contained in diagonal strips of radius $2\sqrt{2}\delta$ centered in the origin. Thus $\vec{c}$ is contained within a tilted square 
of side length $4\sqrt{2}\delta$ centered in the origin. It follows that the maximal Euclidean distance between $\vec{P}_{+}$ or $\vec{P}_{-}$ and $\vec{c}$ is 
$4\sqrt{3}\delta$ (see Fig.~\ref{lemma_4_a}). This yields
\begin{equation}
\label{midpoint_to_intersection}
||\vec{P}_{+}-\vec{c}||, ||\vec{P}_{-}-\vec{c}||\leq 4\sqrt{3}\delta.
\end{equation}

Moreover, since $||\vec{P}_{00}-\vec{P}_{+}||=\frac{1}{2} ||\vec{P}_{00}-\vec{P}_{11}||$ (and similarly for the case with $\vec{P}_{-}$), the minimum value of the denominators in Eq.~\eqref{r_values} occurs when $||\vec{P}_{00}-\vec{P}_{11}||$ is minimized (see Fig.~\ref{lemma_4_b}), that corresponds to the value $2-4\sqrt{2}\delta$. This leads to
\begin{equation}
\begin{aligned}
&||\vec{P}_{+}-\vec{P}_{00}||, ||\vec{P}_{+}-\vec{P}_{11}||, ||\vec{P}_{-}- \vec{P}_{01}||, ||P_{0}-P_{10}||\\
& \quad \geq 1-2\sqrt{2}\delta.
\label{midpoint_to_edge}
\end{aligned}
\end{equation}

Combining Eqs.~\eqref{midpoint_to_intersection} and \eqref{midpoint_to_edge} with Eq.~\eqref{r_values}, it follows that
\begin{equation}
r=\max\{r_+,r_-\}\leq \frac{4\sqrt{3}\delta}{1-2\sqrt{2}\delta}.
\end{equation}

Referring back to Eq.~\eqref{a2a1_expression}, this yields
\begin{equation}
\begin{aligned}
\frac{\alpha_2}{\alpha_1}&=d(P_+,P_-)(1-r)+r\\
&\leq d(P_+,P_-)+r\\
&\leq 2\delta + \frac{4\sqrt{3}\delta}{1-2\sqrt{2}\delta}\\
&=\frac{2(1+2\sqrt{3})\delta-4\sqrt{2}\delta^2}{1-2\sqrt{2}\delta}.
\end{aligned}
\end{equation}

This establishes the first bound. For the second bound, we refer to Eq.~\eqref{upper_bound}:

\begin{equation}
\begin{aligned}
\alpha_3&=\frac{r}{1-r}-d(P_+,P_-)\\
&\leq\frac{r}{1-r}\\
&\leq\frac{\frac{4\sqrt{3}\delta}{1-2\sqrt{2}\delta}}{1-\frac{4\sqrt{3}\delta}{1-2\sqrt{2}\delta}}\\
&=\frac{4\sqrt{3}\delta}{1-2\sqrt{2}\delta-4\sqrt{3}\delta}.
\end{aligned}
\end{equation}

\end{proof}

\onecolumngrid

\newpage 

\twocolumngrid

\section{Proof of Theorem \ref{theorem_POM}}
\label{proof6}

\textit{The optimal success probability in two-bit $\varepsilon$-POM of any operational theory that admits of a parity preserving ontological model satisfies $p(b=x_y)\le \frac{3}{4}+\frac{\varepsilon}{4}.$\\}

\begin{proof}
Suppose $\{P_+,P_-\}$ can be distinguished in a single-shot measurement with probability $\varepsilon$. That is, 
\begin{equation}
\label{P+P-_Distance}
\max_{M=X,Y} \{|\mathcal{P}(k|P_+,M)-\mathcal{P}(k|P_-,M)|\}=\varepsilon.
\end{equation}

This expression can be recast as $\max\{\frac{1}{2}|x_+-x_-|,\frac{1}{2}|y_+-y_-|\}=\varepsilon$. Therefore, by definition we have $d(P_+,P_-)=\varepsilon$. Considering Eq.~\eqref{op_dist}, this means that $s_{\mathcal{O}}^{P_+,P_-}=\frac{1+\varepsilon}{2}.$ Our assumption of parity preservation entails that $d(\mu_+,\mu_-)=\varepsilon$. Thus, Eq.~\eqref{ont_dist} now reads as $s_{\Lambda}^{\mu_+,\mu_-}=\frac{1+\varepsilon}{2}$, which also means that $s_{\mathcal{O}}^{P_+,P_-}=s_{\Lambda}^{\mu_+,\mu_-}$. 
In other words, we have 
\begin{equation}
s_{\Lambda}^{\mu_{00}+\mu_{11},\mu_{01}+\mu_{10}}=s_{\mathcal{O}}^{P_{00}+P_{11},P_{01}+P_{10}}=\frac{1+\varepsilon}{2}.
\end{equation}

It follows from the results of Proposition 2 in \cite{Chaturvedi2020} that 
\begin{equation}
\begin{aligned}
&\frac{1}{2}\left(s_{\mathcal{O}}^{P_{00}+P_{01}, P_{10}+P_{11}}\right)+\frac{1}{2}\left(s_{\mathcal{O}}^{P_{00}+P_{10}, P_{01}+P_{11}}\right)\\
& \quad \leq \frac{1+\frac{1+\varepsilon}{2}}{2}=\frac{3}{4}+\frac{\varepsilon}{4}.
\label{prop_2}
\end{aligned}
\end{equation}

The probability of success is the averaged sum of all possible ways Bob can win, given a randomly chosen measurement $X,Y$ and state $P_{00},P_{01},P_{10},P_{11}$, and an output $0,1$ that Bob guesses based on the outcome of his measurement. This amounts to
\begin{equation}
\begin{aligned}
\label{p_suc}
&p(b=x_y)\\
&=\frac{1}{8} [\mathcal{P}(0|P_{00},X)+\mathcal{P}(0|P_{01},X)\\
& \quad +\mathcal{P}(1|P_{10},X)+\mathcal{P}(1|P_{11},X)\\
& \quad \quad  +\mathcal{P}(0|P_{00},Y)+\mathcal{P}(0|P_{10},Y)\\
& \quad \quad \quad +\mathcal{P}(1|P_{01},Y)+\mathcal{P}(1|P_{11},Y)]\\
&=\frac{1}{2}\left(\frac{ \mathcal{P}(0|\frac{1}{2}P_{00}+\frac{1}{2}P_{01},X)}{2}+\frac{\mathcal{P}(1|\frac{1}{2}P_{10}+\frac{1}{2}P_{11},X)}{2}\right)\\
& \quad +\frac{1}{2}\left(\frac{\mathcal{P}(0|\frac{1}{2}P_{00}+\frac{1}{2}P_{10},Y)}{2}+\frac{\mathcal{P}(1|\frac{1}{2}P_{01}+\frac{1}{2}P_{11},Y)}{2}\right)\\
&=\frac{1}{2}\left(s_{\mathcal{O}}^{P_{00}+P_{01}, P_{10}+P_{11}}\right)+\frac{1}{2}\left(s_{\mathcal{O}}^{P_{00}+P_{10}, P_{01}+P_{11}}\right).
\end{aligned}
\end{equation}

By combining equations \eqref{prop_2} and \eqref{p_suc} we obtain\\ $p(b=x_y)\leq\frac{3}{4}+\frac{\varepsilon}{4}.$

\end{proof}

\section{The case of quantum depolarizing noise}

\label{Appendix_DepolarizingChannel}

\begin{figure}[htbp!]
	\centering
	{\includegraphics[width=.48\textwidth,height=.35\textheight]{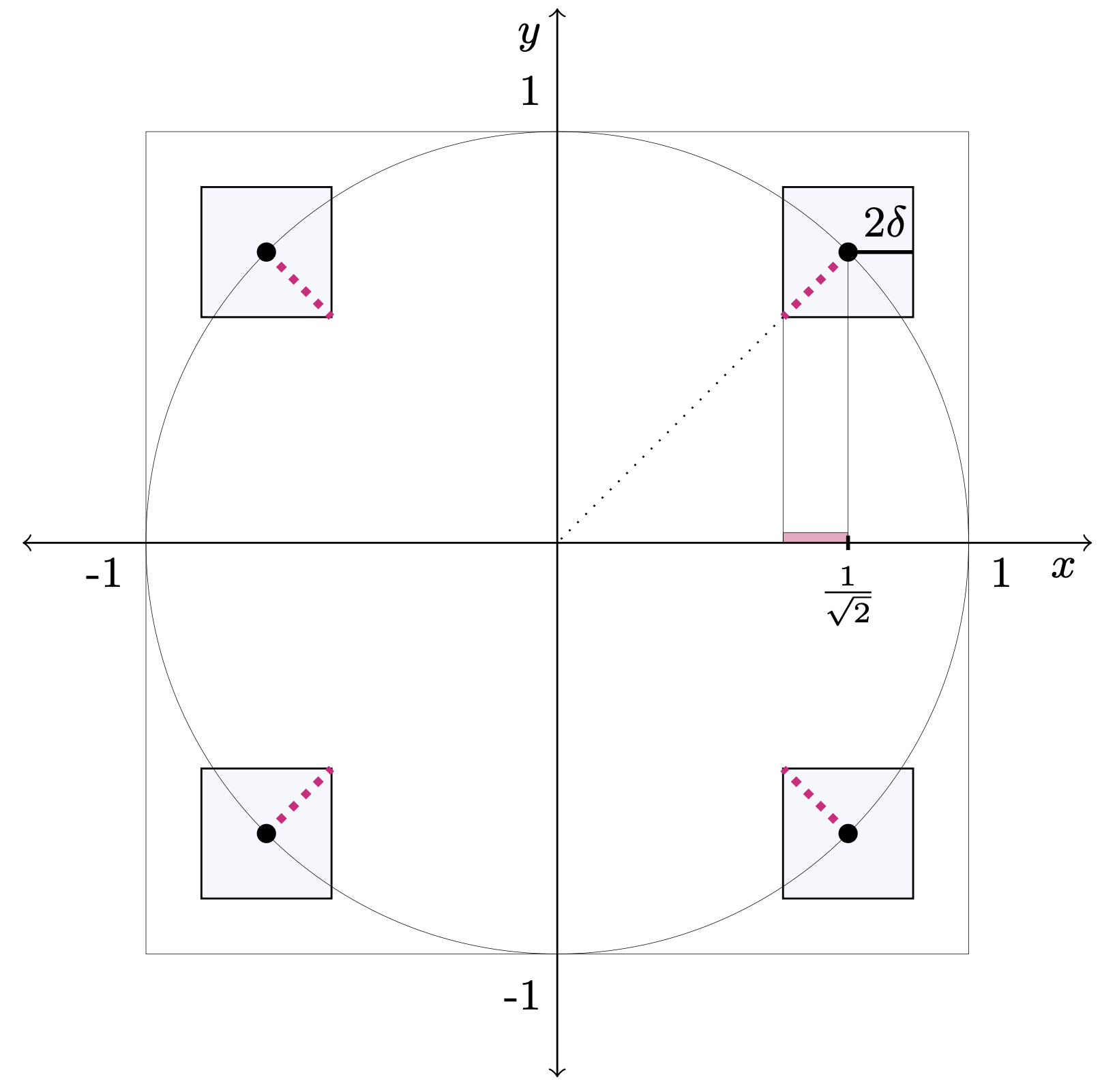}}
\caption{\textbf{Depolarizing noise constrained to a bound of $\delta$}. Ideal preparations are indicated at the center of the shaded squares. Noisy preparations are assumed to lie on the dotted red segments. In accordance with Eq.~\eqref{dist_def_simplest_scenario}, they have an operational distance of \textit{at most} $\delta$ from the ideal ones.}
\label{noise_depolarization}
\end{figure}

In this appendix we focus on the simplest nontrivial scenario in the case where preparations are contained within the unit Bloch disk and the experimental noise is assumed to be modeled by a quantum depolarizing channel. In this case, noisy preparations $P_{ij}$ are mixtures of the ideal preparations $P_{ij}^{id}$ with the completely mixed state $\frac{I}{2}$ (see Fig.~\ref{noise_depolarization}).\\

We begin with an updated bound on Pusey's expression and subsequent threshold for violating preparation noncontextuality based on this bound.

\begin{lemma}
	\label{pusey_lemma_quantum}
	Suppose the preparations $\{P_{ij}\}$ of the simplest scenario satisfy a noise bound $d(\vec{P}_{ij},\vec{P}_{ij}^{id})\leq \delta$, where $\delta$ is the noise parameter and $\{P_{ij}^{id}\}$ are the ideal a priori preparations. Pusey's expression $S(x_{ij},y_{ij})$ of Eq.~\eqref{pusey} satisfies the following lower bound in terms of the noise parameter $\delta$: 
	\begin{equation}
	\label{Pusey_Noise_Inequality_Quantum}
	S(x_{ij},y_{ij})\geq 2\sqrt{2}-2-8\delta + \frac{8\sqrt{2}\delta^2-4\delta}{1-\sqrt{2}\delta},
	\end{equation}
	where $\{x_{ij},y_{ij}\}$ are the coordinates of the preparations $\{P_{ij}\}$.
\end{lemma}

The proof follows that of Lemma \ref{pusey_lemma} in Appendix \ref{proof1}, with the following changes in values: $u_\delta=\frac{1}{\sqrt{2}}$ and $p,q\geq \frac{1-2\sqrt{2}\delta}{2-2\sqrt{2}\delta}$.\\

	


Given that the corresponding preparation noncontextuality inequality is $S(x_{ij},y_{ij})\leq 0$, solving for the right hand side in Eq.~\eqref{Pusey_Noise_Inequality_Quantum} results in a violation threshold of $\delta \approx0.07$, which leads to the following theorem.

\begin{theorem}
	\label{pusey_theorem_quantum}
	If $d(\vec{P}_{ij},\vec{P}_{ij}^{id})\leq 0.07$, then $S(x_{ij},y_{ij})> 0$ and Pusey's preparation noncontextuality inequality is violated. 
\end{theorem}

In the case of quantum depolarizing noise, Lemma \ref{marvian_lemma} and Theorem \ref{marvian_theorem} are left unchanged. This is not the case for Lemma \ref{lemma_bounds}, that provides a different bound than the one in Eq.~\eqref{Upper_Bound_a2a1}. We provide the modified statement below.

\begin{lemma}
	\label{lemma_bounds_quantum}
	Given the functions $\alpha_1$ and $\alpha_2$, if each noisy preparation $P_{ij}$ satisfies $d(\vec{P}_{ij},\vec{P}_{ij}^{id})\leq \delta$, the following upper bound holds:
	\begin{equation}
	\label{Upper_Bound_a2a1_quantum}
	\frac{\alpha_2}{\alpha_1} \leq \delta + \frac{\sqrt{2}\delta}{1-2\sqrt{2}\delta}.
	\end{equation}
\end{lemma}

The proof follows that of lemma \ref{lemma_bounds} in Appendix \ref{proof4}, with the following changes in values: $d(P_+,P_-)\leq \delta$; $\vec{c}=\vec{0}$; $||\vec{P}_{+}-\vec{c}||, ||\vec{P}_{-}-\vec{c}||\leq \sqrt{2}\delta$;  $||\vec{P}_{+}-\vec{P}_{00}||, ||\vec{P}_{+}-\vec{P}_{11}||, ||\vec{P}_{-}- \vec{P}_{01}||, ||P_{0}-P_{10}|| \geq 1-2\sqrt{2}\delta$; and $r\leq \frac{\sqrt{2}\delta}{1-2\sqrt{2}\delta}$. Combining Eq.~\eqref{Upper_Bound_a2a1_quantum} with Eq.~\eqref{implication_1} leads to the following theorem.

\begin{theorem}
	\label{relationship_theorem_noise_quantum}
	Given the simplest scenario with even- and odd-parity preparations $P_+,P_-$ and inaccessible information $C_{\prep}^{\min}$, if each $P_{ij}$ satisfies $d(\vec{P}_{ij},\vec{P}_{ij}^{id})\leq \delta$, then the following implication holds.
	\begin{equation}
	\label{implication_1_noise_quantum}
	C_{\prep}^{\min}> \delta + \frac{\sqrt{2}\delta}{1-2\sqrt{2}\delta} \implies \mathcal{D}_{P_+,P_-}^{\min}>0.
	\end{equation}
\end{theorem}

Equating the right hand sides of inequalities \eqref{Marvian_Inequality} and \eqref{Upper_Bound_a2a1_quantum}, we find the threshold for which Marvian's inequality provides a sufficient lower bound to violate parity preservation as in Eq.~\eqref{implication_1_noise_quantum}, which results in $\delta \approx 0.02$. 
Consequently, the following theorem holds. 

\begin{theorem}
	\label{threshold_pp_quantum}
	If $d(\vec{P}_{ij},\vec{P}_{ij}^{id})\leq 0.02$, then $\mathcal{D}_{P_+,P_-}^{\min}>0$ and parity preservation is violated. 
\end{theorem}

Given that both Marvian's and Pusey's approaches exhibit a violation if $\delta\leq0.07$, we obtain the following noise threshold below which all methods agree in witnessing nonclassicality in the simplest scenario in the case of quantum depolarizing noise. 

\begin{theorem}
	\label{threshold_all_quantum}
	If the noise parameter $\delta$ satisfies $\delta\leq 0.02$, then $S(x_{ij},y_{ij})>0, C_{\prep}^{\min}>0$, and $\mathcal{D}_{P_+,P_-}^{\min}>0$. Therefore, all three criteria of classicality are violated.
\end{theorem}


\end{document}